\documentclass[useAMS]{mn2e}
\usepackage{graphicx}
\usepackage{epsfig}
\usepackage{amssymb}
\usepackage{lscape}
\usepackage{ulem}
\usepackage{txfonts}

\def\c2s{C\,{\sc ii}$^{\star}$}

\def\dfgash2{$\Delta f_{\rm gas, H_2}$}

\title[A definitive merger-AGN connection] {A definitive merger-AGN connection at $z\sim$0 with CFIS: mergers have an excess of AGN \textit{and}  AGN hosts are more frequently disturbed.}

\author[Ellison et al.] {Sara L. Ellison$^1$,  Akshara Viswanathan$^{1,2}$, 
  David R. Patton$^3$,  Connor Bottrell$^1$, \newauthor Alan W. McConnachie$^4$,  Stephen Gwyn$^4$, 
  Jean-Charles Cuillandre$^5$\\ 
$^1$ Department of Physics \& Astronomy, University of Victoria, Finnerty Road, Victoria, British Columbia, 
  V8P 1A1, Canada\\
  $^2$ Anna University, Chennai - 600 025, Tamil Nadu, India\\
  $^3$ Department of Physics and Astronomy, Trent University, 1600 West Bank Drive, Peterborough, ON K9L 0G2, Canada\\
$^4$ NRC Herzberg Astronomy and Astrophysics, 5071 West Saanich Road, Victoria, BC V9E 2E7, Canada\\
  $^5$ AIM, CEA, CNRS, Universit\'e Paris-Saclay, Universit\'e Paris Diderot, Sorbonne Paris Cit\'e, Observatoire de Paris, PSL University, F-91191 Gif-sur-Yvette Cedex, France
}

\begin{document}

\maketitle

\begin{abstract}
  The question of whether galaxy mergers are linked to the triggering of active galactic
  nuclei (AGN) continues to be a topic of considerable debate.  The issue can be broken
  down into two distinct questions:  1) \textit{Can} galaxy mergers trigger AGN?  2)
  Are galaxy mergers the \textit{dominant} AGN triggering mechanism?  A complete picture of
  the AGN-merger connection requires that both of these questions are addressed with the same dataset.
  In previous work, we have shown that galaxy mergers selected from the Sloan Digital Sky
  Survey (SDSS) show an excess of both optically-selected, and mid-IR colour-selected AGN,
  demonstrating that the answer to the first of the above questions is affirmative.
  Here, we use the same optical and mid-IR
  AGN selection to address the second question, by quantifying the frequency
  of morphological disturbances in low surface brightness $r$-band images from the Canada France Imaging
  Survey (CFIS).  Only $\sim$30 per cent of \textit{optical} AGN host galaxies are morphologically
  disturbed, indicating that recent interactions are not the dominant trigger.  However,
  almost 60 per cent of \textit{mid-IR} AGN hosts show
  signs of visual disturbance, indicating that interactions play a more significant role in nuclear feeding.
  Both mid-IR and optically selected AGN have interacting fractions that are a factor of two
  greater than a mass and redshift  matched non-AGN control sample, an excess that increases
  with both AGN luminosity and host galaxy stellar mass.

\end{abstract}

\begin{keywords}
galaxies: active, galaxies: ISM, galaxies: Seyfert
\end{keywords}

\section{Introduction}

It has long been understood that interactions between galaxies can cause a
loss of angular momentum in the gaseous component of the interstellar medium
(ISM), leading to inflows towards the galactic centre (e.g. Barnes \& Hernquist 1991;
but see Gabor et al. 2016 and Blumenthal \& Barnes 2018 for a more
contemporary perspective).  The predicted
effects of these gas inflows, that channel gas inwards from the galactic outskirts,
include central boosts in the star formation rate and dilution of the central
gas phase metallicity (e.g. Montuori et al. 2010; Perez et al. 2011; Torrey et al.
2012; Moreno et al. 2015).  These predictions are well established observationally
(e.g. Kewley et al. 2010; Scudder et al. 2012; Ellison et al. 2013; Patton et al. 2013;
Thorp et al. 2019).

A further prediction of merger-triggered gas inflows is that they can provide fuel
for accretion onto the central supermassive black hole (e.g. Cattaneo et al.
2005; Di Matteo et al. 2005; Capelo et al. 2015, 2017).
However, in contrast to the predictions of the effect of mergers on star formation
and the metallicity of the ISM, the observational connection between mergers and
nuclear accretion has seen much controversy.  There is an extensive literature both
supporting (e.g. Keel et al. 1985; Alonso et al. 2007; Woods \& Geller 2007; Koss et al. 2010, 2012;
Ellison et al. 2011; Silverman et al. 2011; Bessiere et al. 2012; Cotini et al. 2013;
Lackner et al. 2014; Satyapal et al. 2014; Weston et al. 2017; Goulding et al.
2018), and rejecting (Grogin et al. 2005; Li et al. 2008; Gabor et al. 2009;
Reichard et al. 2009; Cisternas et al. 2011; Kocevski et al. 2012; Schawinski et al.
2015; Sabater et al. 2015; Mechtley et al. 2016; Marian et al. 2019) the link between mergers and the
triggering of active galactic nuclei.

There are likely numerous reasons for the apparently conflicting observational results
connecting AGN and mergers.  The first is in the simple definition of what we mean by
`AGN' and how samples are selected.  Accretion onto a supermassive black hole is a
multi-wavelength phenomenon, with radiation emanating from the accretion disk, the dusty
torus, the corona, jets and extended emission line regions, spanning the electromagnetic
spectrum from the X-ray to the radio (e.g. Alexander \& Hickox 2012 for a review).  Not all
AGN are simultaneously characterized by all of these phenomena (e.g. Trouille
\& Barger 2010; Mateos et al. 2013; Mendez et al. 2013; Azadi et al. 2017). Therefore, selecting
AGN with different techniques  leads to AGN (and host galaxies) with different properties
(e.g. Hickox et al. 2009;   Best \& Heckman 2012; Mendez et al. 2013; Gurkan et al. 2014;
Ellison et al. 2016).
The impact of selecting AGN at different wavelengths on the interpretation
of the merger-AGN connection was clearly demonstrated by Satyapal et al. (2014),
who showed that mergers show a greater excess of mid-IR selected AGN, compared with AGN
that are selected on the basis of their optical emission lines.  Several other
observational works have similarly connected mid-IR selected (or simply reddened)
AGN with mergers (Urrutia et al. 2008; Glikman et al. 2015; Fan et al. 2016;
Weston et al. 2017; Goulding et al. 2018; Donley et al. 2018).  These
studies indicate that merger triggered AGN tend to be obscured, and hence more readily
identifiable in the mid-IR.  Simulations have supported this hypothesis, showing that the
high interstellar gas densities and high luminosities associated with merger-induced
accretion lend themselves to identification using mid-IR colours (Blecha et al. 2018).
That obscured AGN are connected with merger activity is also supported by studies of
Compton thick AGN (Kocevski et al. 2015, Ricci et al. 2017; Koss et al. 2016, 2018).
However, whilst both mid-IR
selected and (to a lesser extent) optically selected AGN are more common in mergers than
in isolated galaxies, $z \sim 0$ low excitation radio galaxies appear to be secularly triggered
(Ellison, Patton \& Hickox 2015).  The issue of a `merger-AGN connection' therefore depends
critically how one defines `AGN'.

Secondly, just as AGN are selected using different techniques, so are interacting
galaxies.  Mergers can variously be selected by using spectroscopic galaxy pairs
(e.g. Alonso et al. 2004; Nikolic et al. 2004; Davies et al. 2015),
which can be used effectively even at high redshifts (e.g. Lin et al. 2007;
Wong et al. 2011),
visual classifications (e.g. Darg et al. 2010; Kartaltepe et al. 2015),
automated parametric morphology metrics
(Lotz et al. 2004; Pawlik et al. 2016), or machine learning techniques
(e.g. Ackermann et al. 2018; Walmsley et al. 2019).  All of these methods have pros
and cons.  For example, in the era of large spectroscopic surveys, pair catalogs
can be generated easily and contain large numbers, yet they often suffer
from incompleteness (e.g. Patton \& Atfield 2008).  The human eye is a very
effective search tool, but this method is subjective and time consuming. Parametric
morphologies are quick to run, and reproducible, but are incomplete and likely
sensitive to different stages in the merger sequence and different
galaxy/interaction properties (e.g. Nevin et al. 2019).
Machine learning is both reproducible, efficient and reaches beyond the parametric
limitations of quantitative morphologies, but can be limited by the training
set.  Even for a given merger selection technique, identification will
depend on the characteristics of the parent sample, such as depth and wavelength.

A third complication in the definition of AGN samples, and their connection with mergers,
is the possibility of a luminosity dependence.  Some of the earliest evidence for a
connection between nuclear accretion and mergers came from observations of the very
brightest members of the AGN family: quasars (Heckman et al. 1984;
Hutchings et al. 1984, 1988; Stockton \& MacKenty 1987).  The most luminous radio-loud AGN
are also dominated by mergers (Ramos-Almeida et al. 2011; 2012) and there is evidence
that merger-induced
AGN activity leads to more luminous AGN (e.g. Ellison et al. 2011; Liu et al. 2012;
Satyapal et al. 2014; Alonso et al. 2018).
However, there remains debate about whether there is a systematic dependence on the likelihood
of a merger origin and the AGN luminosity (e.g. Treister et al. 2012, versus Hewlett
et al. 2017; Villforth et al. 2017).

Finally, it is important to recognize that the issue of a merger-AGN connection is actually
composed of two separate questions.  The first is whether mergers \textit{can} trigger
AGN and the second is whether mergers are the \textit{dominant} AGN trigger.
The answer to the first question appears to have been definitively (at least at low $z$)
answered by the clear statistical excess of AGN in merger (or close pair) selected samples
(Woods \& Geller 2007; Ellison et al. 2011, 2013, 2015; Khabiboulline et al. 2014;
Satyapal et al. 2014; Weston et al. 2017). However,
the fact that the frequency of morphological disturbances in the host galaxies of AGN is
similar to that of non-AGN control galaxies has been used to argue that mergers are not the dominant
AGN trigger (Gabor et al. 2009; Cisternas et al. 2011; Kocevski et al. 2012; 
Mechtley et al. 2016). Simulations have suggested the same, with mergers contributing
to, but not necessarily dominating, the fuelling pathways of AGN (e.g. Draper \& Ballantyne
2012; Bellovary et al. 2013; Menci et al. 2014; Steinborn et al. 2018).

Taken at face value, the above observations would seem to argue for a picture in
which mergers are a possible, but not dominant mechanism for AGN triggering.  However,
as argued above, care must be taken when synthesizing the results from studies using
different AGN selection techniques, in different luminosity regimes and also at different
redshifts.  Indeed, most of the studies that find no excess for morphological disturbances
in AGN host galaxies are for X-ray selected AGN at $z>1$, in contrast to the studies that find
enhanced optical and mid-IR selected AGN fractions in mergers at $z \sim 0$. A more
fair assessment of the merger-AGN
connection requires using a single dataset to address both questions: 1) can mergers trigger
AGN?  2) are most AGN triggered by mergers?

In previous work, we have used samples of close galaxy pairs, and recently coalesced
post-mergers selected from the Sloan Digital Sky Survey (SDSS) to demonstrate that
mergers \textit{can} trigger optically selected AGN (Ellison et al. 2011, 2013; Satyapal
et al. 2014), hence providing an affirmative
answer to the first of the two questions above.  In this paper, we present a complementary
analysis in which we aim to answer the second question by quantifying how widespread AGN
triggering is in the same SDSS sample.  Therefore, whereas our
previous work started with a merger sample and counted AGN, here we begin with an AGN
sample and count the frequency of morphologically disturbed galaxies using deep $r$-band
imaging from the Canada-France Imaging Survey (CFIS).

The paper is laid out as follows.  In Section \ref{sec_methods} we describe
our methods, including description of the AGN selection (both optical and mid-IR),
compilation of the control samples, the CFIS imaging dataset and morphological
classification schemes.  Our results are presented in Section \ref{sec_vis_results}
with discussion in Section \ref{sec_discuss} and a summary in Section \ref{sec_summary}.
We adopt a cosmology in which H$_0$=70 km/s/Mpc, $\Omega_M$=0.3, $\Omega_\Lambda$=0.7.

\section{Methods}\label{sec_methods}

\subsection{The Canada-France Imaging Survey}

\begin{figure*}
	\includegraphics[width=18cm]{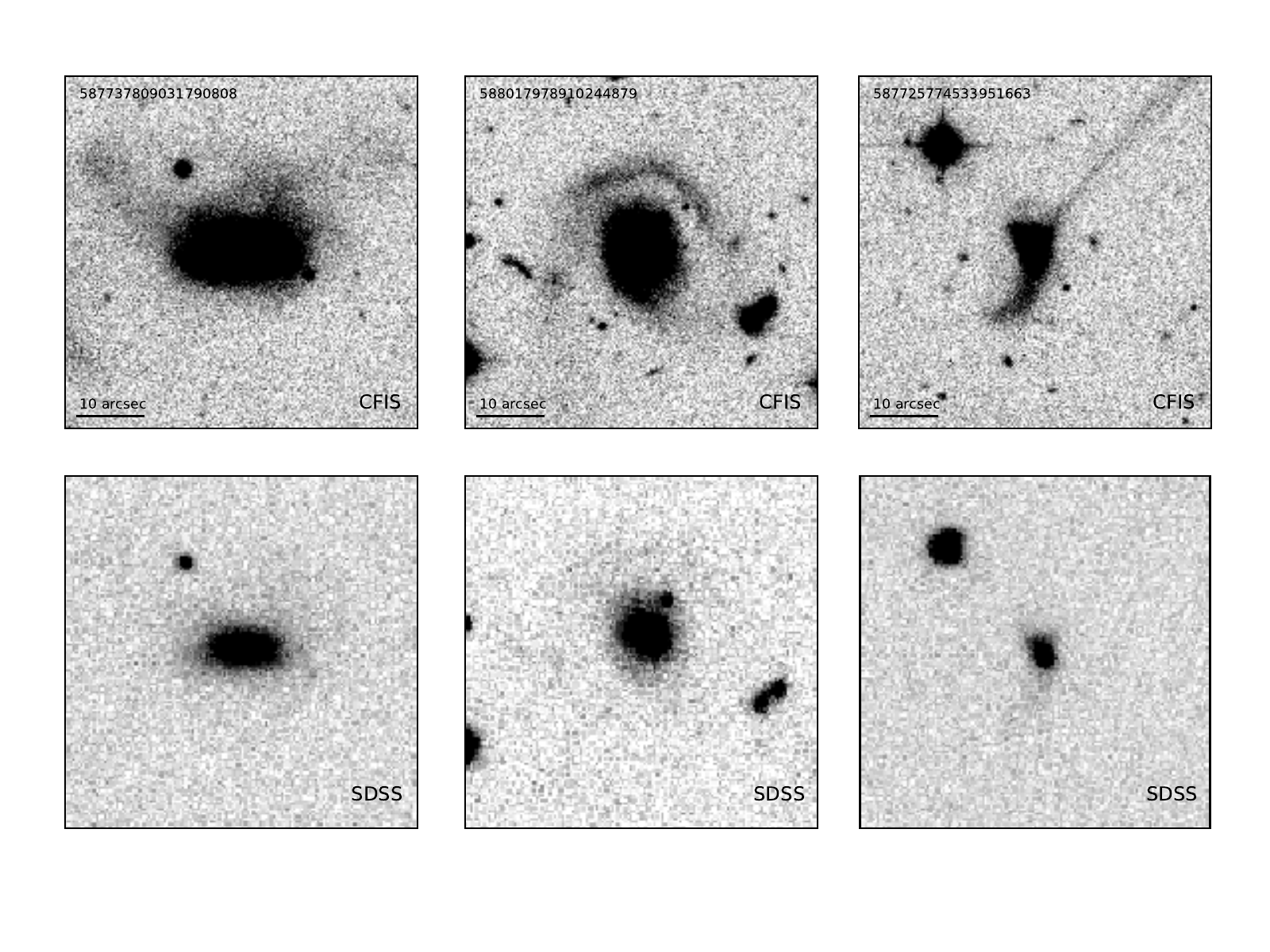}
        \caption{Top row: CFIS $r$-band images for 3 galaxies.  The image
          scale is 0.18 arcsec/pixel.  The SDSS DR7
          objID is given in the top left corner of each panel.  Bottom row:
          SDSS $r$-band images for the same three galaxies.  The image scale is 0.4
        arcsec/pixel.}
    \label{cutouts_fig}
\end{figure*}

\begin{figure}
	\includegraphics[width=\columnwidth]{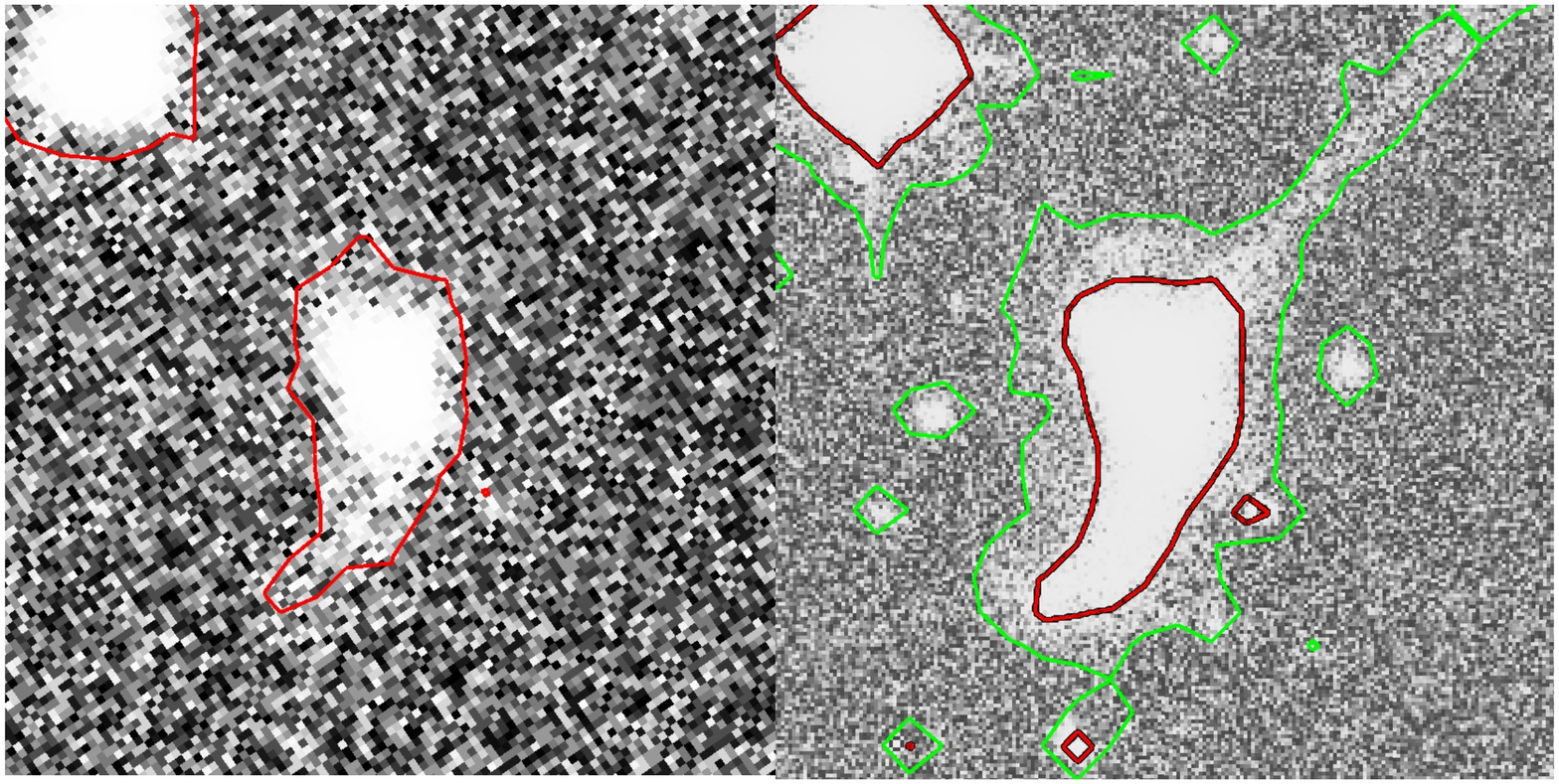}
        \caption{Left panel: Red contours show the limiting isophote for
          the SDSS $r$-band image ($r$ = 24.4 mag arcsec$^{-2}$) for
          galaxy objID=587725774533951663 (right column of Fig. \ref{cutouts_fig}).
          Right panel: Red contours show the same isophote for the CFIS $r$-band image.
          Green contours show the limiting isophote for the CFIS $r$-band image
          ($r$ = 27.1 mag arcsec$^{-2}$), demonstrating the the CFIS image extends
          almost 3 mag arcsec$^{-2}$ deeper than the SDSS image.}
    \label{jc_iso}
\end{figure}

CFIS is a Large Program, jointly led by Canada and France,
currently on-going at the 3.6-metre Canada-France-Hawaii
Telescope on Maunakea.  The survey is conducted using the MegaCam wide-field optical
imager, whose field of view covers a full square degree on the sky.
The survey consists of observations in the $r$ and $u$ bands 
with 5 $\sigma$ point-source depths of 
24.85 and 24.4 mag in the $r$ and $u$ bands, respectively.
During the lifetime of the survey (nominally 2017-2021) CFIS-$r$ will
image $\sim$ 5000 sq. deg. above a declination of 30 degrees of which $\sim$ 3300
sq. deg. overlaps with the SDSS-Baryon Oscillation Spectroscopic Survey (BOSS)
footprint.  The $u$-band observations will cover an even larger area -- 10,000
sq. deg above a declination of 0 (for first results from the $u$-band, see Ibata et al. 2017).

The CFIS observing strategy consists of three visits of one exposure each, with a third of the camera field of view offset in between each of the three exposures. This approach optimizes astrometric and photometric absolute calibration with respect to varying observing conditions, while ensuring any area on the survey footprint is covered by at least two exposures (the vast majority being naturally covered by three exposures). The large dithering allows a nocturnal sky background map to be derived on short time windows (30 minutes) and ensures a proper internal flattening of the background to restore the true flat nocturnal sky which in turns enables low surface brightness studies (Cuillandre et al., 2019, in prep).

The MegaCam images are processed at CFHT by the Elixir pipeline (Magnier \& Cuillandre 2004) and delivered to CADC where they are calibrated against the Pan-STARRS PS1 survey (Magnier et al. 2016) for photometry, leading to a 0.4 percent absolute calibration, and Gaia DR2 (Brown et al. 2016) for astrometry with residuals just over 20 mas. CADC's MegaPipe pipeline's end products consist of optimized stacks and object catalogs.

As of May 2018, approximately 1900 sq. degrees of sky coverage had been obtained
and processed and combined into 0.5 $\times$ 0.5 degree tiles in the $r$-band;
these are the CFIS data that are used
for the present study.  Approximately 110,000 galaxies with spectroscopic redshifts
in the Main Galaxy Sample in the SDSS Data Release 7 (DR7)
are covered by these $r$-band tiles.

In Fig. \ref{cutouts_fig} we demonstrate the quality of the CFIS $r$-band images
by comparing the CFIS image (top row) with SDSS $r$-band image (bottom row)
for 3 galaxies.   A further selection of
mergers identified in CFIS imaging is presented in the Appendix (Fig. \ref{multi_plots_fig}).
The superior quality of the CFIS $r$-band images compared
to the SDSS DR7 images is evident, a combination of the greater depth and
better image quality of the former over the latter.  
Whereas the median $r$-band seeing
for the CFIS data is 0.65 arcsec (with a dispersion of only 0.1 arcsec),
it is only $\sim$ 1.5 arcsec for the
SDSS images.  The pixel scale is also much finer for MegaCam than the SDSS CCDs:
0.18 arcsec compared to 0.4 arcsec.  The postage stamps in Fig. \ref{cutouts_fig}
demonstrate that the CFIS images are able to reveal faint tidal features that
are invisible on the SDSS images.  

In order to provide a rough comparison between the surface brightness limits in
SDSS and CFIS, we compared isophotes for the galaxy in the right
hand column of Fig. \ref{cutouts_fig} (SDSS objID=587725774533951663).
The limiting isophote on the SDSS image
(red outline in left panel of Fig \ref{jc_iso}) corresponds to 24.4 mag
arcsec$^{-2}$.  For the CFIS image (right panel of Fig. \ref{jc_iso}) the limiting
isophote (green contour) is almost 3 magnitudes deeper: 27.1 mag arcsec$^{-2}$.

\subsection{AGN sample selection}

AGN are a multi-wavelength phenomena and the various selection techniques used
to identify them target different structural components, such as the accretion
disk, corona, jets or dusty torus.  There is compelling observational evidence
that selecting AGN in different wavelength regimes identifies different populations
(e.g. Hickox et al. 2009; Mendez et al. 2013; Azadi et al. 2017).
Although there is overlap between some of these detection techniques,
some AGN are only identified at certain wavelengths, due either to the absence of
certain components (e.g. radio jets or gas reservoirs), inclination effects
or other observational biases.  Importantly for this study, it has been argued that
not all `flavours' of AGN are likely to be triggered by mergers (e.g. Ellison et al.
2015), and that the fraction of AGN associated with mergers likely depends on
the selection procedure (Satyapal et al. 2014; Blecha et al. 2018).  We therefore
identify AGN in our sample using two complementary techniques, using both optical
emission line diagnostics and mid-IR colours.

\subsubsection{Optical AGN sample}

The optical AGN sample is based on galaxies in the SDSS DR7, as classified by the
Kewley et al. (2001) scheme, which identifies galaxies strongly dominated by AGN emission.
We require that a S/N $>$ 5 is achieved in all four of the emission lines on the AGN
diagnostic diagram.  This criterion largely excludes Low Ionization Nuclear Emission
Line Regions (LINERs) from our sample, which is desirable given that many of these
sources are not associated with AGN (e.g. Yan \& Blanton 2012; Belfiore et al. 2016).
Another advantage of using the Kewley et al. (2001) AGN classification is that the luminosity of
the [OIII] emission line can be used as an indicator of AGN luminosity.  However,
we do not attempt to convert the [OIII] luminosity to a bolometric luminosity - the
bolometric correction is both uncertain and would only result in a systematic shift
to all our values.  The [OIII] luminosity can therefore still be used as an indicator
of relative AGN power within our sample.

\begin{figure}
	\includegraphics[width=\columnwidth]{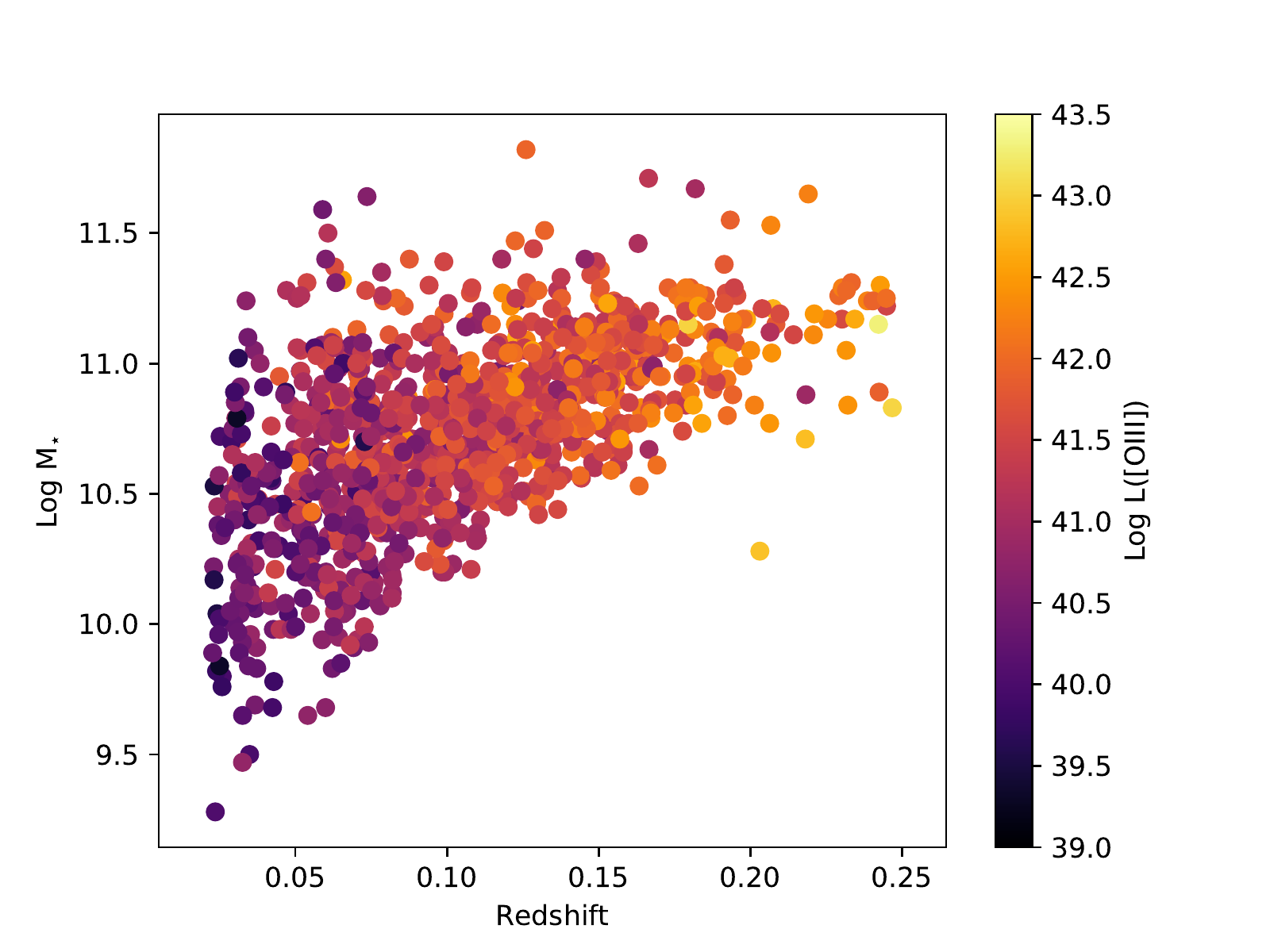}
        \caption{Stellar mass and redshift distribution of the
          optical AGN sample, with points colour-coded
        by the luminosity (log erg/s) of the $\lambda$ 5007 [OIII] line.}
    \label{opt_sample_fig}
\end{figure}

Taken together, the Kewley et al. (2001) selection scheme and high S/N
requirement yield a robust sample of galaxies whose emission lines are dominated
by AGN.  There are a total of 8419 AGN in the SDSS DR7 that fulfill these criteria.
The positions of these 8419 DR7 AGN were cross-matched with the CFIS $r$-band
survey coverage, as of May 2018.  The requirement of overlap with CFIS reduced the
sample to 1269 AGN.  However, upon visual inspection, 145 of these galaxies
coincided with incomplete tiles, chip gaps or artefacts, leaving a sample of 1124 optical AGN.
This relatively high fraction of `missing' coverage is the result of the
on-going nature of the survey, where not all tiles have yet had their observations
completed.  The distribution of stellar mass, redshift and the luminosity of the
$\lambda$ 5007 [OIII] line (taken from the MPA/JHU catalog, e.g. Kauffmann et al. 2003)
for the optical AGN sample is presented in Fig. \ref{opt_sample_fig}.

\subsubsection{Mid-IR AGN sample}

AGN can be identified from red mid-IR colours that result from radiation
from the heated dust torus  (e.g. Stern et al. 2005, 2012).  We cross-match the Wide-Field Infrared
Survey Explorer (WISE; Wright et al. 2010) to positions of galaxies in the
SDSS DR7, requiring a match within 6 arcseconds (the full width at half maximum of the WISE point
spread function in the W1 band).  Various mid-IR colour cuts have been
  proposed in the literature, both for WISE and other mission band passes (e.g. Donley
  et al. 2007; Assef et al. 2013).
  The exact choice depends on several considerations, including the redshift range, balance of
  completeness versus purity, and the extent to which star formation may contribute
  (e.g. Hainline et al. 2016; Blecha et al. 2018).
As was the case for the optical AGN
sample described in the previous sub-section, for the purposes of our
work it is preferable to select galaxies that are dominated by AGN
emission in the mid-IR, with relatively little contribution from star formation or
other ionizing processes.  We therefore require a S/N $>$ 5 in
both the $W1$ and $W2$ bands and impose a colour cut of $W1 - W2 > 0.8$
(Stern et al. 2012) in order to identify the mid-IR AGN sample.  There are 300 mid-IR AGN
so selected.

\begin{figure}
	\includegraphics[width=\columnwidth]{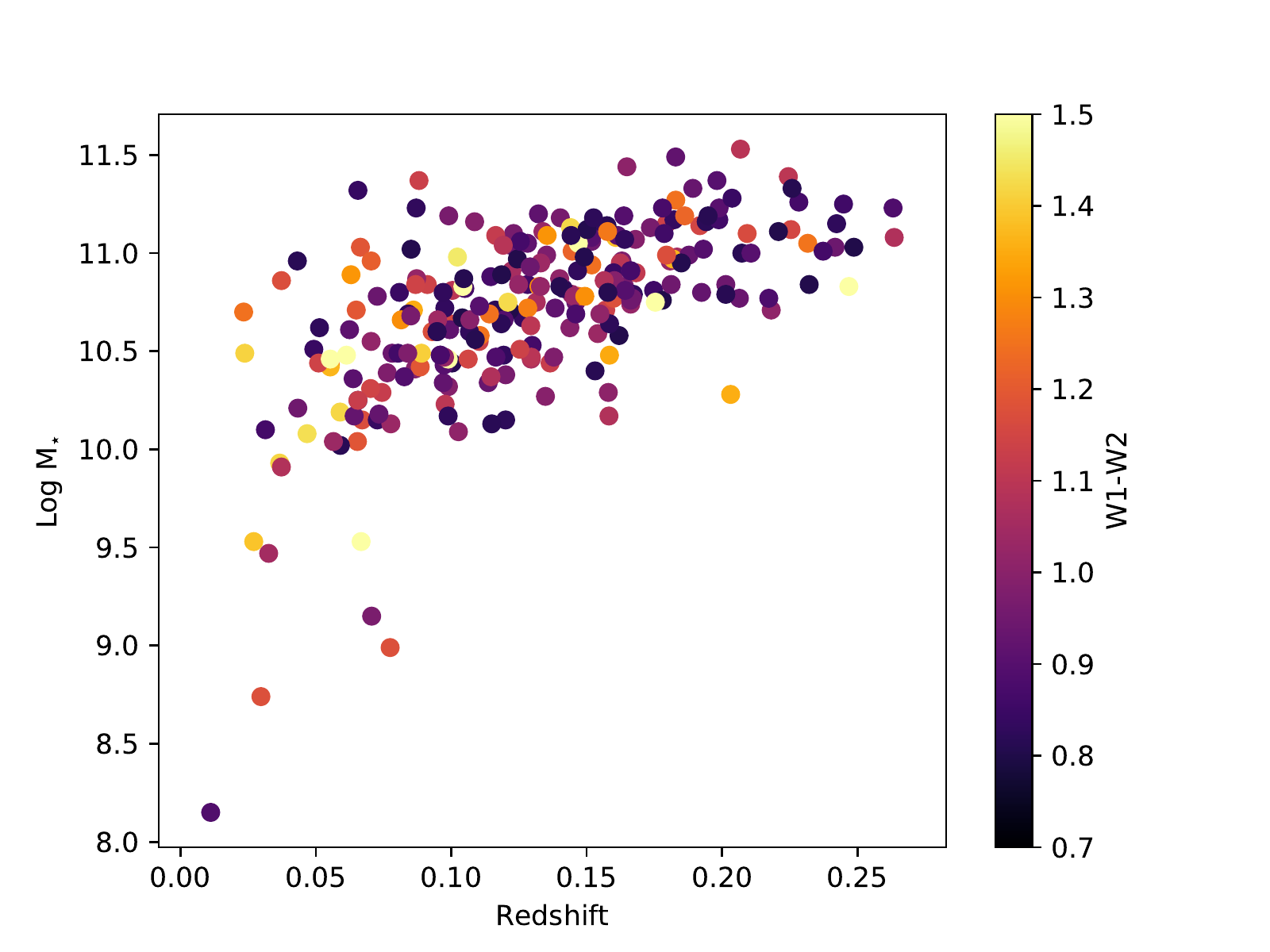}
        \caption{Stellar mass and redshift distribution of the mid-IR
          AGN sample, with points colour-coded by $W1-W2$ colour. Note that
          the $y$-axis has been expanded, compared with Fig. \ref{opt_sample_fig},
        in order to capture the handful of low mass galaxies in the WISE sample.}
    \label{wise_sample_fig}
\end{figure}

As was the case for the optical sample described in the previous sub-section,
some of the mid-IR selected AGN fall on chip gaps,
incomplete tiles or
are otherwise affected by artifacts that render them unsuitable for
visual classification.  The final WISE AGN sample therefore contains
254 galaxies with CFIS coverage in the $r$-band.  The distribution of
stellar mass, redshift and the $W1-W2$ colour
for the mid-IR AGN sample is presented in Fig. \ref{wise_sample_fig}.

\subsection{Control matching}

The control matching procedure is based on a `pool' from which controls that are matched
in stellar mass and redshift may be drawn.  Control matching is done separately for the
optical and mid-IR selected AGN samples, such that each of the two AGN samples has
its own pool.

\subsubsection{Optical AGN control pool}

For the optically selected AGN sample, the control pool is taken to be any galaxy
in the DR7 not classified to be an AGN according to the Kauffmann et al. (2003)
emission line criterion.  Note that although we used the Kewley et al. (2001)
criterion to identify AGN, using the Kauffmann et al. (2003) demarcation
for selecting non-AGN control galaxies conservatively removes from the control
pool galaxies that have both star-forming and AGN contributions (commonly referred
to as composites).  The control pool for the optical AGN therefore contains
galaxies classified as either star-forming according to Kauffmann et al. (2003),
or with no emission line classification (i.e. they can be passive galaxies).
Moreover, we relax the S/N criterion for the selection of the control pool;
the S/N $>$ 5 threshold for AGN was chosen to aid the purity of the AGN sample
against LINER contributions.  The selection of non-AGN does not have this concern.
The control pool is therefore composed of galaxies with S/N$>$3 that are
classified as star-forming by Kauffmann et al. (2003), as well galaxies with
emission lines below this S/N threshold (essentially, non-star-forming galaxies).
There are approximately 429,000 control pool galaxies in the SDSS DR7 that
fulfill these criteria, of which 68,223 are in CFIS data coverage.

\subsubsection{Mid-IR AGN control pool}

For the mid-IR selected AGN sample, the control pool is taken to be any galaxy
in the DR7 with $W1-W2 < 0.5$.  Note that, as was the case for the optically selected
sample, we have put a buffer between our definition of AGN (selected to
have $W1-W2 > 0.8$) and control galaxies.  This is again to yield cleaner samples;
whilst $W1-W2 > 0.8$ is generally accepted as a colour cut that identifies AGN
dominated colours, obscured AGN may have bluer colours in the range $0.5 <W1-W2<0.8$
(e.g. Blecha et al. 2018).  The $W1-W2<$0.5 criterion therefore provides a cleaner
sample of non-AGN galaxies for the control pool.
There are approximately 308,000 mid-IR control pool galaxies in the SDSS DR7, of which
50,994 are in the current CFIS data coverage.

\subsubsection{Matching procedure}

For both AGN (optical and mid-IR) samples, the single best match control galaxy for each
AGN was selected from its respective control pool by simultaneously minimizing the
difference in stellar mass and redshift.  Although we do not expect much redshift evolution
of physical properties within the range of our sample (Figs. \ref{opt_sample_fig}
and \ref{wise_sample_fig}), the redshift matching is
important to ensure that the effective surface brightness imaging depth is consistent
between the AGN and their controls.   We required that the difference in log
stellar mass between the AGN and the control could be no greater than 0.1 dex and the
redshift must match within $\pm$0.005.    If no control galaxy was found within
these tolerances, the AGN was removed from the sample.  However, in practice,
we found that every AGN had an adequate control galaxy available in the pool, and that the
typical difference between the stellar masses and redshifts was less than 0.02 dex and
0.001 respectively.

\subsection{Visual morphological classification}\label{vis_morph_sec}

Visual classifications of CFIS $r$-band cutouts
were carried out blindly, i.e. with no knowledge of whether the galaxy was an AGN or a control
galaxy.  The cutout was examined interactively, so that the classifier could change
the dynamic range and scaling of the image in order to fully scrutinize the presence
of morphological features.  Typically, such features either manifested as faint extended
tails/shells, or asymmetries (including double nuclei) within the half light radius of the galaxy.
These two circumstances require very different dynamic ranges in image scaling in
order to identify them, making manual manipulation a vital part of the classification
process.  The fiducial morphology flags used for our analysis (unless
otherwise stated) are those by author Ellison (SLE), but we repeated
our analysis with co-author Patton's (DRP) classifications to check that the results were
qualitatively similar. 

Many previous works have devised morphological classification schemes (e.g.
Nair \& Abraham 2010; Kocevski et al. 2012; Kartaltepe et al. 2015; Larson et al. 2016).
In this work, we opted to keep the process simple, since the fundamental question
we seek to address is whether the galaxy in question is (or has been) recently
part of an interaction.  We therefore do not attempt to break down the classification
scheme into many different merger stages.  Instead, we first assess simply whether a
given galaxy shows \textit{convincing} signs of morphological disturbance, such as tails or shells.
If such features are present, the classifier next assesses whether there is
an obvious companion that causes these disturbances.  If a clear perturbing
companion is present, the galaxy is classified as an \textit{interacting pair}.
The identification of a perturbing companion requires more than just proximity;
there must be convincing evidence of a tidal interaction between the two galaxies.
If no companion is present, or if there are nearby galaxies but no obvious
tidal connection, the galaxy is labeled as a \textit{post-merger}.  In this
scheme, a disturbed galaxy with a companion will be classified as a post-merger
if that companion is not obviously participating in the perturbation.  The
interacting pair and post-merger categories together represent all galaxies
that show clear morphological signs of some on-going, or past, interaction.
We therefore refer to these categories combined as \textit{interacting} galaxies.

\begin{figure}
	\includegraphics[width=\columnwidth]{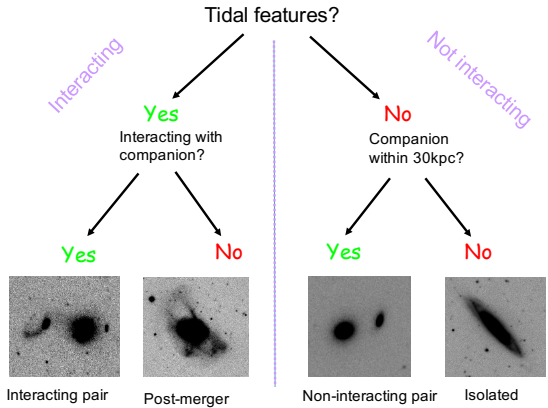}
        \caption{Summary of the visual classification decision tree.  Galaxies are
          classified into four types, based on whether or not
          they are morphologically disturbed, and whether they have a companion.
        Example CFIS images of each class are shown.}
    \label{flowchart_fig}
\end{figure}

\begin{table*}
\begin{tabular}{lcccccc} \hline  \hline
 & Total  & Isolated & non-interacting pairs & interacting pairs & post-mergers & interacting (pairs + PM)  \\ \hline
\# AGN &  1124 & 714 &63  & 86 & 261 & 347 \\
\# non-AGN & 1124 & 880 &  64 & 33 & 147 & 180 \\ \hline \hline
\end{tabular}
\caption{Number of galaxies in each morphological class for the optically selected
  AGN sample and its non-AGN control.}
\label{opt_agn_tab}
\end{table*}

If there is no sign of morphological disturbance, we search the CFIS image for possible
companion galaxies within 30 kpc (projected).  If no companion is found, we label the galaxy
as \textit{isolated}.  If a companion is located, the galaxy is categorized as
a \textit{non-interacting pair}.  Although non-interacting pairs show no signs
of visible disturbance, they may be galaxies on their first approach (i.e.
pre-first pericentre), or galaxies whose internal properties or orbits are less
likely to produce tidal features.  A graphical summary of the classification
process is shown in Fig \ref{flowchart_fig}.

Visual classification of galaxy morphologies is an inherently subjective and
imperfect process.  There will always be borderline cases where the classifier
struggles to decide whether or not a given galaxy should be considered
disturbed or not.  Moreover, as is evident from Fig \ref{cutouts_fig},
the visibility of tidal features is clearly dependent on the depth of the
imaging.  Indeed, the paradigm of hierarchical growth implies that \textit{all}
massive galaxies are merger remnants and very deep imaging shows that faint
tidal structures are extremely common, even in `normal' galaxies (e.g. Tal et al.
2009). It is also well known that the visibility timescale of merger features
is highly variable, depending on galaxy properties, galaxy orientation,
time since pericentre and orbital parameters
(e.g. Di Matteo et al. 2007; Lotz et al. 2010a,b; Ji \& Peirani 2014).
Therefore, \textit{any} dataset and classification technique should
not be trusted to yield the absolute frequency of interactions, but the comparison
to a control sample can be used to more robustly measure an excess.
In classifying interacting galaxies as those with \textit{convincing}
tidal features/morphological disturbances, we are erring on the side of
purity, at the expense of completeness in our merger categories. 

We also note two subtle differences between the classification
procedure employed here and those commonly used.
First, in contrast to the common practice of combining visual
classifications from multiple expert astronomers (e.g. Kartaltepe et al.
2015; Larson et al. 2016) or citizen scientists (e.g. Lintott et al. 2008) into a single flag
for a given galaxy, in our work, we consider the classification from
each of the two classifiers (SLE and DRP) separately, repeating all of our analysis with
each person's merger flags.  Whilst individual classifiers may differ in
their opinion of a given galaxy with a different perceived level of
asymmetry required by them in order for a galaxy to receive one of the
`interacting' flags, as long as they have classified \textit{consistently}
throughout the AGN and control sample, a comparison in morphologies of the
two is still valid.  Second, one of the benefits of our interactive (i.e.
scaleable) classification process,
over one that produces prescribed image cutouts is that automated
image scaling is often not ideal for the identification of merger features.
Many of the mergers identified in our sample would have been missed in the CFIS
images without the ability to re-scale the contrast interactively.

\section{Results}\label{sec_vis_results}

\subsection{The optical AGN-merger connection}

\begin{figure}
	\includegraphics[width=\columnwidth]{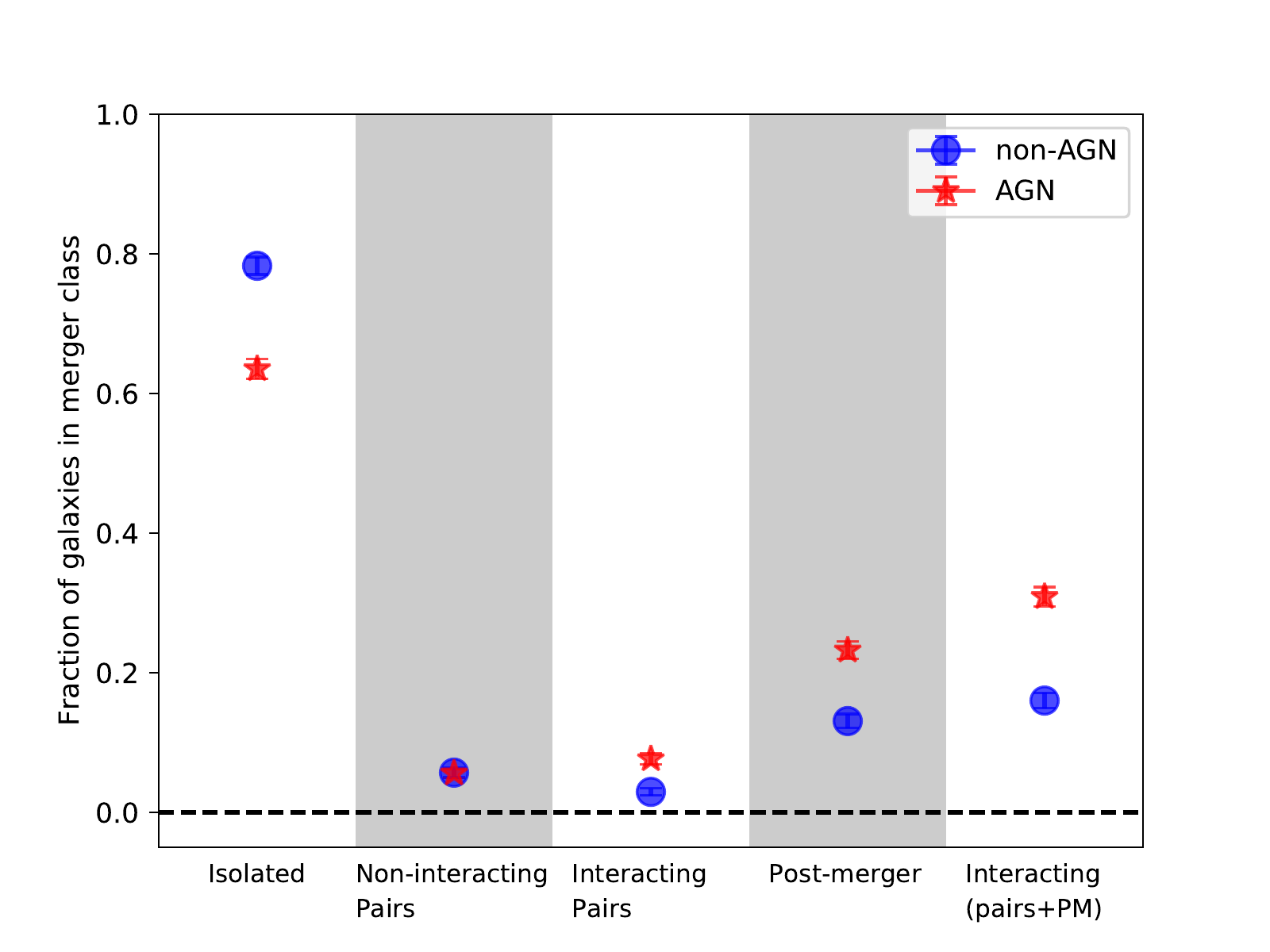}
        \caption{Fraction of optically selected AGN (star symbols) and non-AGN controls
          (circles) in each of
          the morphology classes, as assessed by visual inspection.  Errors
            are the standard errors as computed from binomial statistics. The `interacting'
          category in the right-most column includes interacting pairs and post-mergers and
          hence encompasses all galaxies with discernable tidal features/disturbances.
          Approximately 60 per cent of AGN are in isolated galaxies, indicating that the
          dominant AGN triggering mechanism is not by a recent merger.  However, the enhanced
          fraction of AGN, compared with non-AGN controls, in interacting pairs and post-mergers
        indicates that mergers do contribute statistically to the AGN population.}
    \label{AGN_frac_fig}
\end{figure}

We begin our analysis by computing the fraction of optical AGN in each
of the four basic morphology classes described in Sec. \ref{vis_morph_sec}:
isolated, non-interacting pair, interacting pair and post-merger.  These
statistics are listed in Table \ref{opt_agn_tab} and shown graphically
in Fig. \ref{AGN_frac_fig} as red stars, where error bars
are the standard errors as computed from binomial statistics.  In addition, a
fifth `interacting' category combines the
interacting pairs and post-merger categories.  For comparison,
the fraction of the control sample in each classification category is shown
as blue circles.  From Fig. \ref{AGN_frac_fig} we see that almost
80 per cent of non-AGN galaxies are classified as isolated, i.e. single
galaxies lacking any obvious morphological disturbance.  This is in
stark contrast to the optical AGN sample in which only 63 per cent
have been classified as isolated. Fig. \ref{AGN_frac_fig} therefore
convincingly demonstrates that optical AGN hosts are less likely to
be in morphologically undisturbed and isolated systems.  However,
since up to 63 per cent of AGN are
in isolated galaxies\footnote{This is an upper limit on the undisturbed fraction
  since deeper imaging or less conservative classification could classify
  weaker features as tidal distortions.}, it appears that \textit{recent interactions are not
the dominant trigger for nuclear accretion in optically selected
AGN}.

Turning next to the non-interacting pair category. 
Figure \ref{AGN_frac_fig} shows that both AGN and non-AGN have similar
fractions of non-interacting pairs, of only $\sim$ 5 per cent.  
Several scenarios could lead to a galaxy being placed in the non-interacting
pair category.  First, chance projection of a background/foreground galaxy;
AGN and non-AGN should be similarly susceptible to such `contamination'.
Second, the two galaxies may be destined to interact, but have not yet
completed their first pericentre passage.  Since the gas inflows that feed
the AGN are not expected to occur pre-pericentre, there should be no
causal connection between AGN triggering, so we would again  expect
similar fractions of non-interacting pairs in the AGN and non-AGN
samples.  Finally, the pair
may have completed a pericentric passage, but on an orbit, or with
galaxy orientations that
are not conducive to tidal disruption.  In this case, the gas flows
should be inefficient, again leading to ineffective AGN triggering.
Therefore, the similarity in the non-interacting pair fractions
between AGN and non-AGN is not surprising.

The deficit in isolated AGN hosts (compared with non-AGN hosts)
is made up by higher fractions of AGN in
the two interacting categories.  The interacting pairs and post-mergers
represent 8 and 23 per cent (respectively) of the AGN sample, i.e.
a total of 31 per cent of optical AGN hosts have been
classified as exhibiting convincing morphological disturbances. We
stress that this should be considered as a lower limit to the true fraction of
disturbed AGN hosts, due both to the conservative nature of our classification
and also the limited sensitivity of the CFIS imaging.  In contrast,
only 16 per cent of non-AGN are classified as interacting.  \textit{Our results
therefore show that optical AGN are twice as likely to be found in
interacting galaxies than non-AGN, but that the majority of optical
AGN are in non-interacting galaxies at $z \sim 0$.}

\begin{figure}
	\includegraphics[width=\columnwidth]{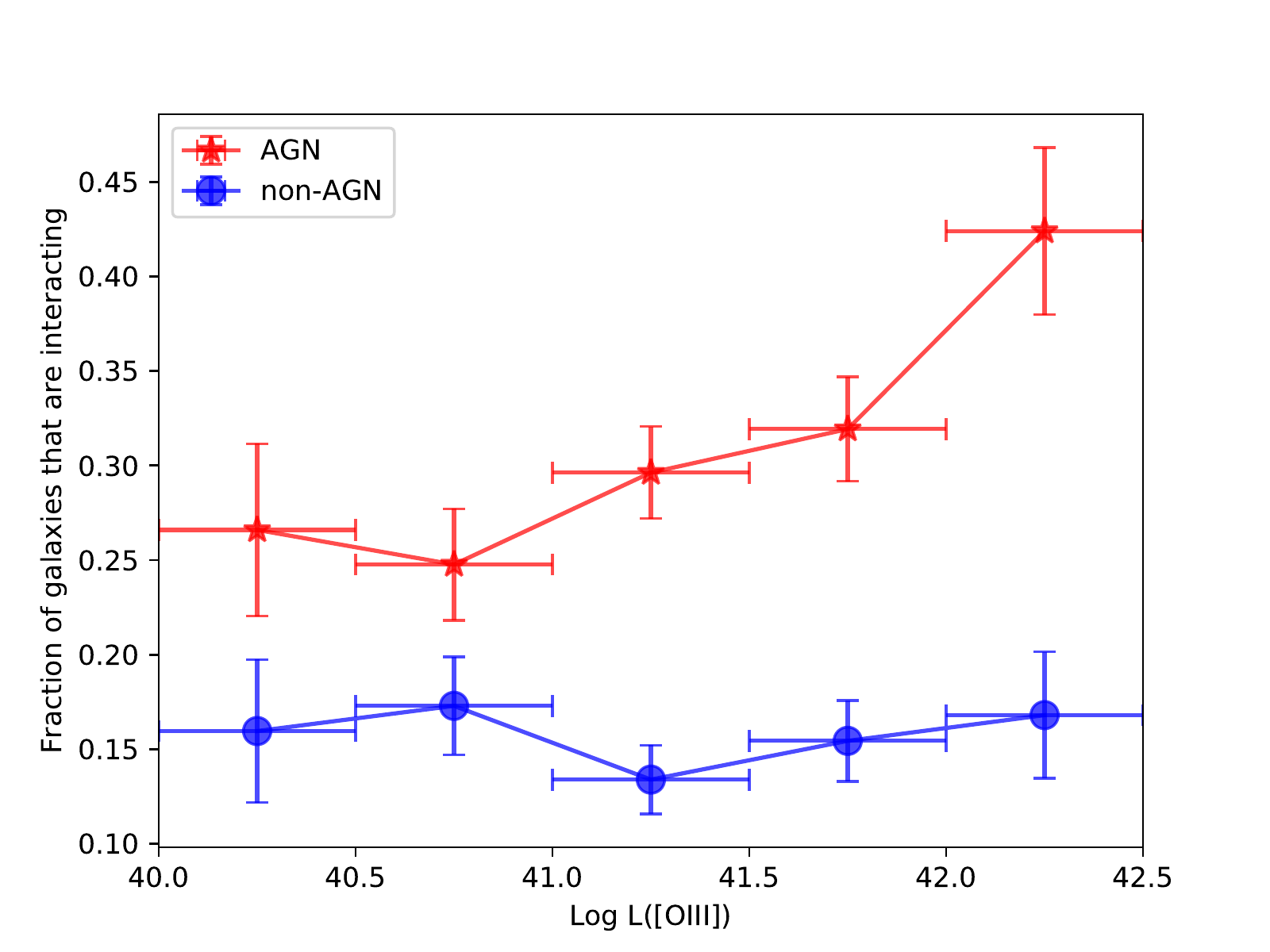}
	\includegraphics[width=\columnwidth]{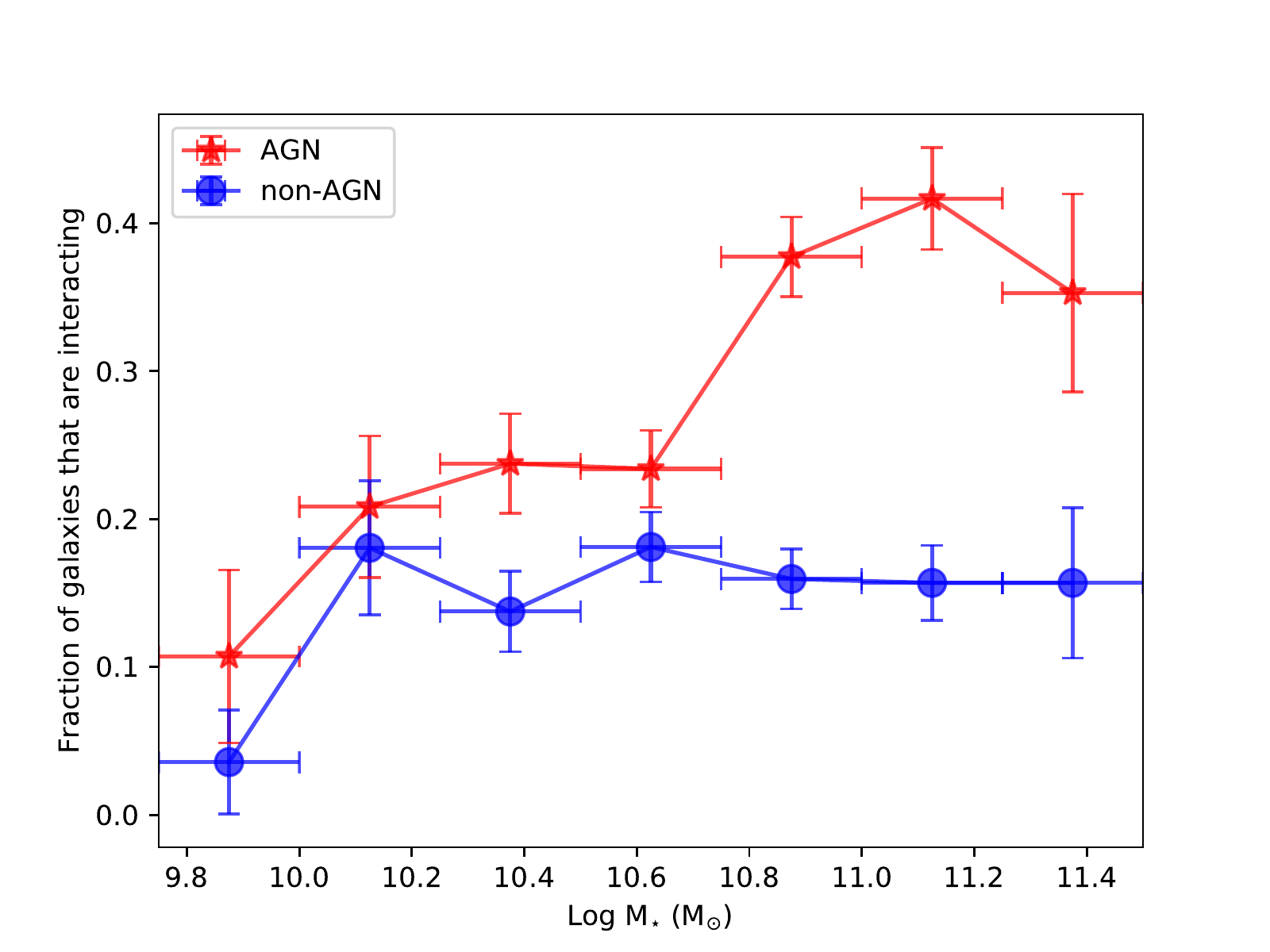}
        \caption{The fraction of optically selected AGN (red stars) and non-AGN control galaxies
          (blue circles) that are classified as interacting
          (interacting pairs plus post-mergers) as a function of the [OIII] line
          luminosity (top panel) and galaxy stellar mass (lower panel). Errors
            in the x-dimension indicate the bin width and errors in the y dimension
            are the standard errors computed from binomial statistics. }
    \label{AGN_ex_LO3_fig}
\end{figure}

Fig. \ref{AGN_frac_fig} indicates that up to $\sim$ 60 per cent of optical AGN in the
local universe may be triggered by processes other than galaxy interactions.
Although luminous quasars frequently
seem to be associated with mergers, there remains debate about whether or not
there is a correlation between AGN luminosity and merger incidence (e.g.
Treister et al. 2012 vs. Villforth et al. 2014).
In the upper panel of Fig. \ref{AGN_ex_LO3_fig} we investigate whether the more
luminous AGN in our sample are more likely to be triggered by mergers.
Since we have selected optical AGN based
on the Kewley et al. (2001) diagnostic, the [OIII] luminosity is a good indicator of
AGN luminosity, with minimal contamination from star formation.

In bins of log L([OIII]),
we count the number of AGN that are classified as interacting (interacting pairs
and post-mergers).  We similarly count the number of controls matched to
the AGN in each  log L([OIII]) bin that are interacting.
Therefore, although the controls do not have an
(AGN relevant) L([OIII]) luminosity themselves, we can still assess the AGN morphologies
relative to the controls to which they were matched.

In the upper panel
of Fig. \ref{AGN_ex_LO3_fig} we plot the fraction of interacting AGN as a function of
L([OIII]) as red stars.  For comparison, we plot as blue circles the fraction of
interacting controls (non-AGN) in the L([OIII]) bins that correspond to their matched AGN.  That is,
the fraction of interacting AGN in a given L([OIII]) bin is compared to the interacting
fraction of those same galaxies' controls.  The upper panel of Fig. \ref{AGN_ex_LO3_fig}
shows that the interacting fraction of AGN exceeds that of the non-AGN sample at
all values of L([OIII]).  However, whereas the fraction of interacting controls
remains constant with log L([OIII]), the fraction of interacting AGN
steadily increases from $\sim$ 25 per cent at
log L([OIII]) $\sim$ 40.5 erg/s, to $\sim$ 45 per cent at log L([OIII]) $\sim$ 42.5 erg/s.
The visual impression of an increasing interaction fraction in AGN with L([OIII])
is confirmed with a Pearson correlation test, which yields a correlation
coefficient of 0.89.  In order to account for the uncertainties on each data
point, fractions in the upper panel of Fig. \ref{AGN_ex_LO3_fig}
are re-sampled 10,000 times drawing at random from a gaussian distribution
with standard deviation set by the error bars on each point.  The correlation
coefficient is found to be $>$ 0.75 in 65 per cent of re-samplings and $>$ 0.65
in 80 per cent of re-samplings.  

Although it has long been predicted that mergers will lead to high AGN luminosities
  due to the high accretion rates thus triggered (e.g. Di Matteo et al. 2005),
  the dependence of merger triggering on AGN
  luminosity has been challenged by some recent
  simulations.  For example, although Steinborn et al. (2018) find
  an excess of mergers amongst AGN selected from their cosmological simulations (in agreement
  with our findings), they find no
  correlation of the merger fraction and AGN luminosity at $z \sim$0, in contrast with our
  observational results.  Moreover, the fraction of merging AGN at  $z \sim$0 in these simulations
  is only $\sim$ 10 per cent, in clear contradiction with our results.  Although
  subtleties such as methods for selecting AGN, and identifying merger features affect
  these absolute numbers, it appears that merger
  triggered AGN are much more common in reality than in some simulations.

In the lower panel of Fig. \ref{AGN_ex_LO3_fig} we investigate whether the fraction
of morphologically disturbed optical AGN depends on the stellar mass of the host galaxy.
Once again, the AGN and control samples are shown as red stars and blue circles
respectively.  As we saw for the dependence on L([OIII]) in the upper panel, the control
sample shows no dependence on disturbed fraction with stellar mass.  The fraction
of interacting AGN again exceeds that of the controls at all stellar masses, but
increases by a factor of two, from approximately 20 to 40 per cent from log (M$_{\star}$/M$_{\odot}$)
$\sim$ 10.4 to 11.2.  The visual impression of an increasing interaction fraction with stellar mass
is confirmed with a Pearson correlation test, which yields a correlation
coefficient of 0.91.  The data points in the lower panel of Fig. \ref{AGN_ex_LO3_fig}
are re-sampled 10,000 times drawing at random from a gaussian distribution
with standard deviation set by the error bars on each point.  The correlation
coefficient is found to be $>$ 0.75 in 79 per cent of re-samplings and $>$ 0.65
in 90 per cent of re-samplings.  \textit{We conclude that morphologically
  disturbed fraction of optical AGN host galaxies is significantly greater for higher
  AGN luminosities and larger host galaxy masses.}

The dependence of merger-related AGN triggering on host galaxy stellar
mass might be caused by the correlation of inflow efficiencies on internal
galaxy properties.  For example, high gas fractions, larger gas disks and
higher bulge fractions have all been demonstrated to yield lower rates of star
formation rate enhancement and/or gas infall rates (e.g. Di Matteo et al. 2007;
Cox et al. 2008; Scudder et al. 2015; Fensch et al. 2017; Blumenthal \&
Barnes 2018).  The inefficiency of gas inflows in the presence of a bulge
works counter to the trend seen in the lower panel of Fig \ref{AGN_ex_LO3_fig},
since the typically higher bulge fractions found in high mass galaxies might
be expected to suppress the nuclear fuelling in high mass interactions.
Conversely, the low gas fractions in high mass galaxies could
potentially explain the trend for increasing interaction frequency in
AGN hosted by high mass galaxies.

We also investigated whether there is
  any trend of specific L([OIII]) (L([OIII])/M$_{\star}$) with merger fraction, but found no
  correlation with the fraction of AGN that are interacting.

\subsection{The mid-IR AGN-merger connection}

\begin{figure}
	\includegraphics[width=\columnwidth]{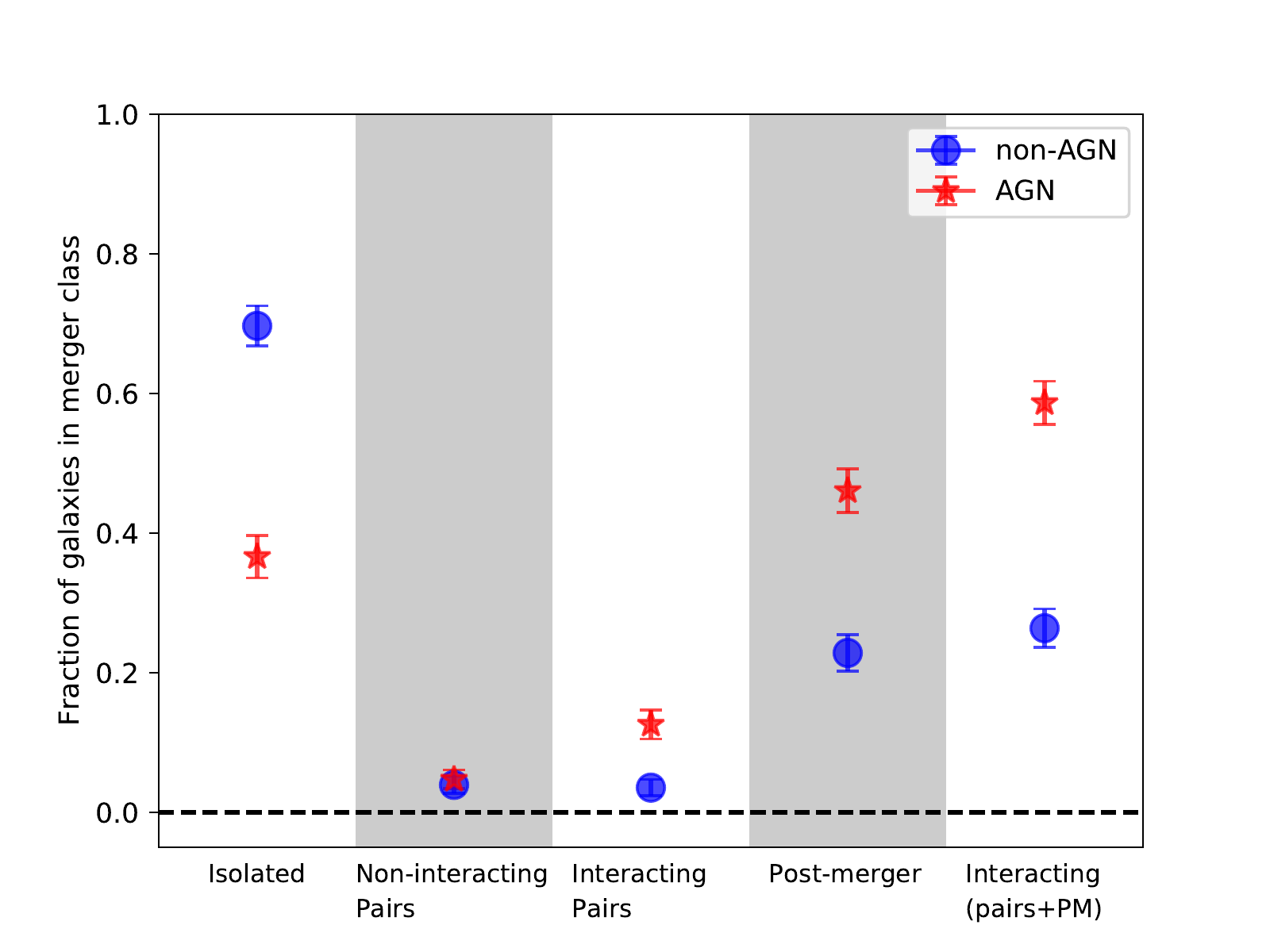}
        \caption{Fraction of mid-IR selected AGN (star symbols) and non-AGN controls (circles) in each of
          the morphology classes, as assessed by visual inspection.  Errors are the standard errors as computed from binomial statistics.  The `interacting'
          category in the right-most column includes interacting pairs and post-mergers and
          hence encompasses all galaxies with discernable tidal features/disturbances.
          At least 59 per cent of AGN are in interacting galaxies, indicating that the
          mergers are potentially the most prevalent mechanism for triggering mid-IR
          AGN.  }
    \label{wise_AGN_frac_fig}
\end{figure}

\begin{table*}
\begin{tabular}{lcccccc} \hline  \hline
 & Total  & Isolated & non-interacting pairs & interacting pairs & post-mergers & interacting (pairs + PM)  \\ \hline
\# AGN &  254 & 93 & 12  & 32 & 117 & 149 \\
\# non-AGN & 254 & 117 & 10 & 8 & 58 & 66 \\ \hline \hline
\end{tabular}
\caption{Number of galaxies in each morphological class for the mid-IR selected
  AGN sample and its non-AGN control.}
\label{ir_agn_tab}
\end{table*}

In Table \ref{ir_agn_tab} and Figure \ref{wise_AGN_frac_fig} we show the
fraction of mid-IR AGN and control
galaxies in the different morphological classes.  The results for mid-IR
AGN in Figure \ref{wise_AGN_frac_fig}
shows several of the same qualitative characteristics that we saw for
optical AGN in Figure \ref{AGN_frac_fig}: a lower fraction of isolated AGN hosts
compared with controls, similar fractions of non-interacting pairs and
elevated fractions of disturbed morphologies amongst AGN hosts.  However,
the interacting fractions are quite different between the optical and mid-IR
selected AGN samples.
Although we re-iterate our caution that absolute fractions depend on
image depth and the subjectivity of the classifier, a relative comparison
between samples is justified.
Whereas 31 per cent of optical AGN hosts were classified as interacting,
the interacting fraction is 59 per cent for mid-IR selected AGN.  The majority
of mid-IR selected AGN are therefore in morphologically disturbed hosts.
Whereas we concluded from Fig. \ref{AGN_frac_fig} that most of
the optically selected AGN had not experienced a recent interaction,
Figure \ref{wise_AGN_frac_fig} shows that \textit{interactions may
  be the dominant triggering mechanism for low $z$ mid-IR selected AGN.}

\begin{figure}
	\includegraphics[width=\columnwidth]{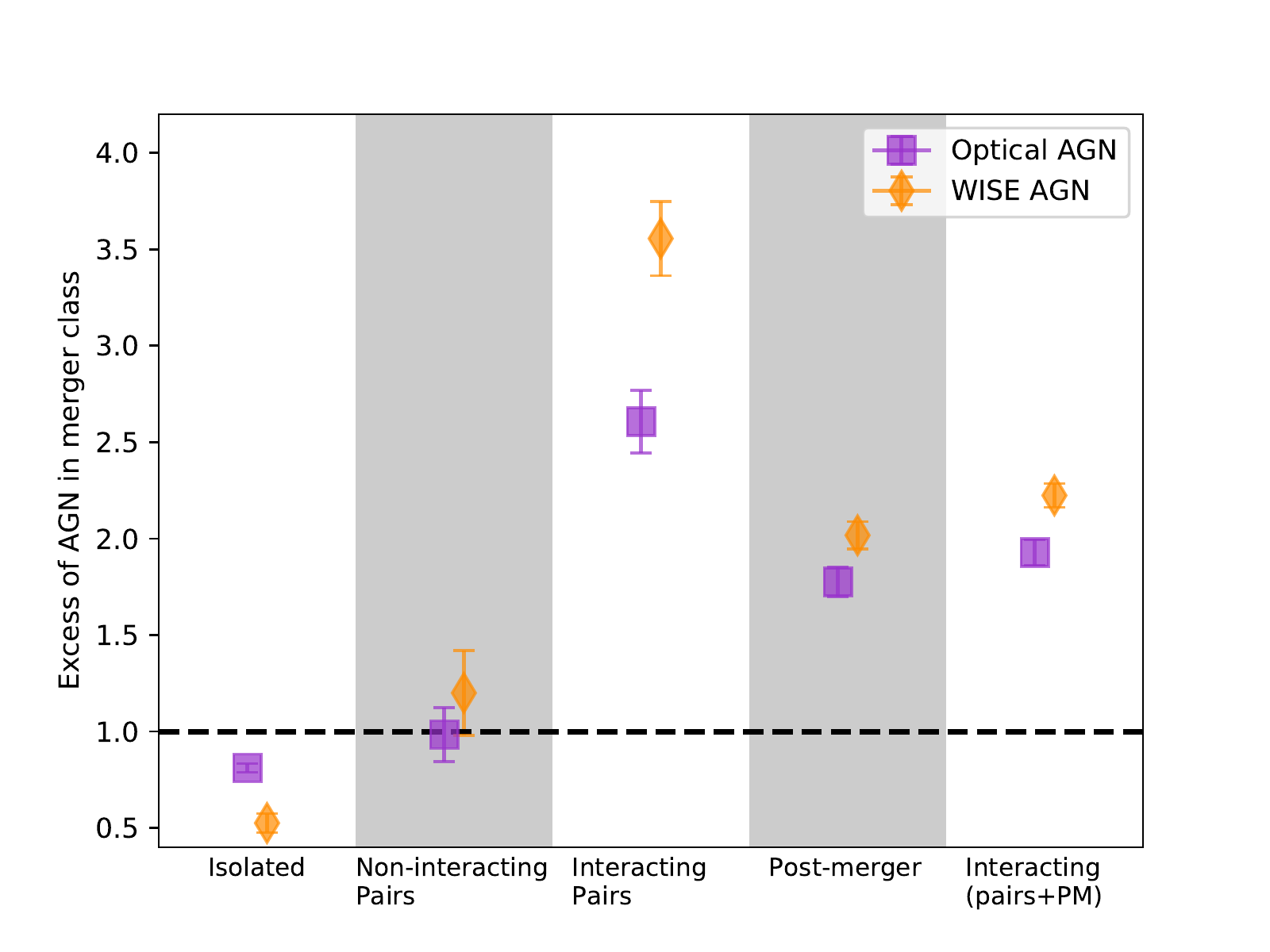}
        \caption{The excess of AGN in each morphology class, i.e. the number of AGN
          divided by the number of controls in each category.  Errors are the standard errors as computed from binomial statistics.  The horizontal dashed
          line shows a ratio of one, as would be expected if AGN and non-AGN are equally
          represented in each morphology class.  AGN are statistically over-represented
          in both interacting pairs and post-mergers, by factors of $\sim$ 2--3.5, with greater enhancements
          seen for the mid-IR selection. The `interacting'
          category in the right-most column includes interacting pairs and post-mergers and
        hence encompasses all galaxies with discernable tidal features/disturbances.}
    \label{AGN_ex_fig}
\end{figure}

In Fig. \ref{AGN_ex_fig} we compare the ratio of AGN-to-control fraction
in each morphological class, for both the optical (square symbols) and
mid-IR (diamonds) AGN samples.  \textit{Broadly speaking, both optical and mid-IR selected AGN are
  approximately twice as likely to be interacting.}  However, the excess
  of interactions is greater in the mid-IR selected AGN, compared with the optically
  selected AGN.  Although this effect is modest for the post-mergers, it is highly
  significant for interacting pairs.  Optically selected AGN are a factor of 2.5 more
  likely to be in an interacting pair, compared to a non-AGN, whereas WISE selected
  AGN show an interacting pair excess of 3.5.  These results add further support
  to an enhanced connection between obscured AGN and mergers (e.g. Satyapal et al.
  2015; Koss et al. 2016, 2018; Weston et al. 2018).

We note that all of the above results are qualitatively similar
  if we use the classifications from the secondary classifier (DRP), instead
  of those of SLE.  Although only approximately 50 per cent of
  galaxies had agreement on absolute merger category, the relative excess between
  AGN and controls is very similar between classifiers.
  This consistency highlights that regardless of the different perceived classifications
  for about half of the sample, as long as the classifier is consistent, then the
same results are achieved.

\section{Discussion}\label{sec_discuss}

Our previous work on galaxy pairs and post-mergers has found
an elevated incidence of optical and mid-IR selected AGN compared to non-interacting
galaxies (Ellison et al. 2011; Ellison et al. 2013; Satyapal
et al. 2013; Khabiboulline et al. 2014; Ellison 2015),
demonstrating that mergers are capable of triggering nuclear activity.
Conversely, many works that have searched for morphological signatures of
disturbances, or companions, have found a similar fraction of interacting
galaxies amongst AGN and control galaxies (e.g. Grogin et al. 2005;
Cisternas et al. 2011; Kocevski et al. 2012).  How can these
apparently conflicting results be reconciled?

One possible explanation that is often cited
is that whilst mergers are capable of triggering AGN,
most AGN are triggered by other mechanisms.  Hence, experiments
that search for an excess of merger features in an AGN sample
may come to a different conclusion to studies that measure more
AGN in a merger sample.  Whilst this scenario offers a reconciliation
of the current set of observational findings, it is predicated on combining samples
from very different redshifts and AGN selected with different methods and
with different luminosities.  In the work presented here, we assess
the morphologies of AGN and control samples selected analogously to
our previous work on mergers, in order to provide a fair assessment of
this scenario.  We have shown that the morphological disturbances are
seen (at least) twice as often in AGN hosts, and that this
excess is seen for both optical and mid-IR selected AGN  (Fig \ref{AGN_ex_fig}).
Therefore, at z $\sim$ 0 the AGN-merger connection is clearly detectable
regardless of whether the experiment searches for an excess of
AGN in mergers (Ellison et al. 2011; Ellison et al. 2013; Satyapal
et al. 2013; Khabiboulline et al. 2014; Ellison, Patton \& Hickox 2015),
or whether we search for an excess of morphological disturbances in
AGN host galaxies (this work).  Moreover, we have shown that the
\textit{majority} of mid-IR selected AGN are interacting galaxies,
and that the excess of interacting hosts is greater for mid-IR
selected AGN than for optically selected AGN.

What are the alternative explanations for the conflicting results
published in the literature?  An obvious possibility is that most
of the papers that find no merger-AGN connection are at higher
redshift, implicating either redshift evolution of
triggering mechanisms or the observational limitations of
detecting tidal features at high $z$.

For reference, the observational bias of bolometric surface brightness dimming
leads to 3 magnitudes reduction in sensitivity from $z=0$ to $z=1$.
Although many high $z$ studies use high quality
Hubble Space Telescope data, many of the interaction features seen
in our low $z$ sample are very faint (e.g. Fig \ref{cutouts_fig}).
In order to test the impact of image quality, we repeat the
morphological classification of the WISE AGN sample (and its control)
using SDSS $r$-band images, which is both shallower (by approximately
3 mag arcsec $^{-2}$) and of
poorer spatial resolution than CFIS.

\begin{figure}
	\includegraphics[width=\columnwidth]{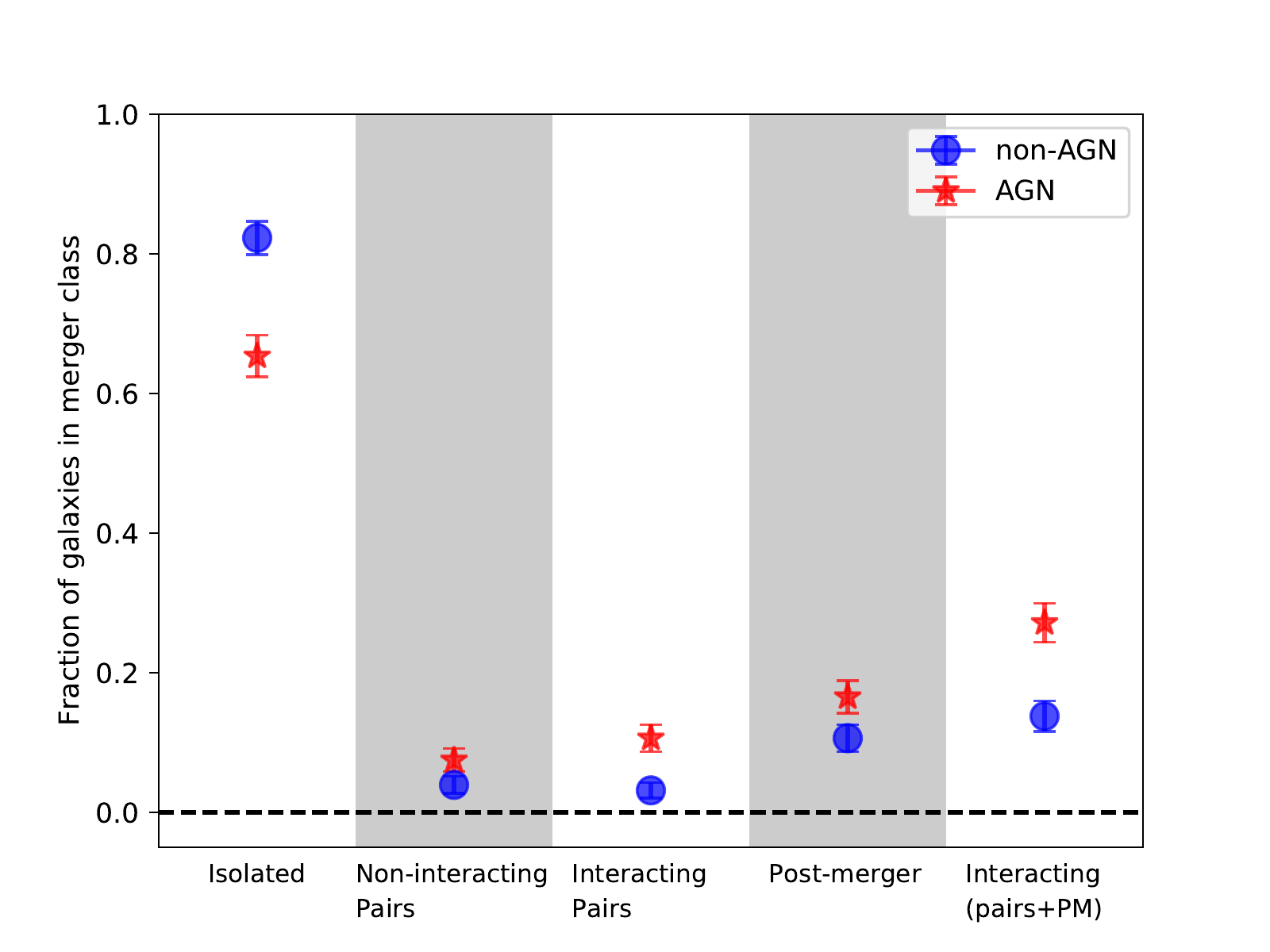}
        \caption{Fraction of mid-IR selected AGN (star symbols) and non-AGN controls (circles) in each of
          the morphology classes, as assessed by visual inspection based on the SDSS
          $r$-band imaging.  Errors are the standard errors as computed from binomial statistics.}  The `interacting'
          category in the right-most column includes interacting pairs and post-mergers and
          hence encompasses all galaxies with discernable tidal features/disturbances.
          In contrast to the classifications based on CFIS images (Fig \ref{wise_AGN_frac_fig})
          the majority of WISE selected AGN appear to be undisturbed in SDSS imaging.  
    \label{wise_sdss_AGN_frac_fig}
\end{figure}

The SDSS classification results are shown in
Fig \ref{wise_sdss_AGN_frac_fig}, which can be compared to Fig.
\ref{wise_AGN_frac_fig} for the same WISE AGN and control sample,
but using CFIS imaging.  There is a dramatic reduction in the fraction
of WISE AGN that are classified as interacting: whereas $\sim$ 60 per cent
of mid-IR selected AGN show morphological disturbances in CFIS imaging,
only $\sim$ 25 per cent have detectable features in the SDSS imaging.
The poorer quality imaging therefore would have definitely affected our
ability to conclude that mergers are dominant amongst WISE AGN.  In their
comparison of SDSS Legacy imaging with the deeper Stripe 82 data, Bottrell
et al. (2019) have recently demonstrated that quantitative morphologies,
as measured by the asymmetry parameter, are also affected by image depth.
However, the controls are also affected by the degraded image quality and
Fig. \ref{wise_sdss_AGN_frac_fig} shows that morphological disturbances
are still more common in WISE AGN than their controls.

\begin{figure}
	\includegraphics[width=\columnwidth]{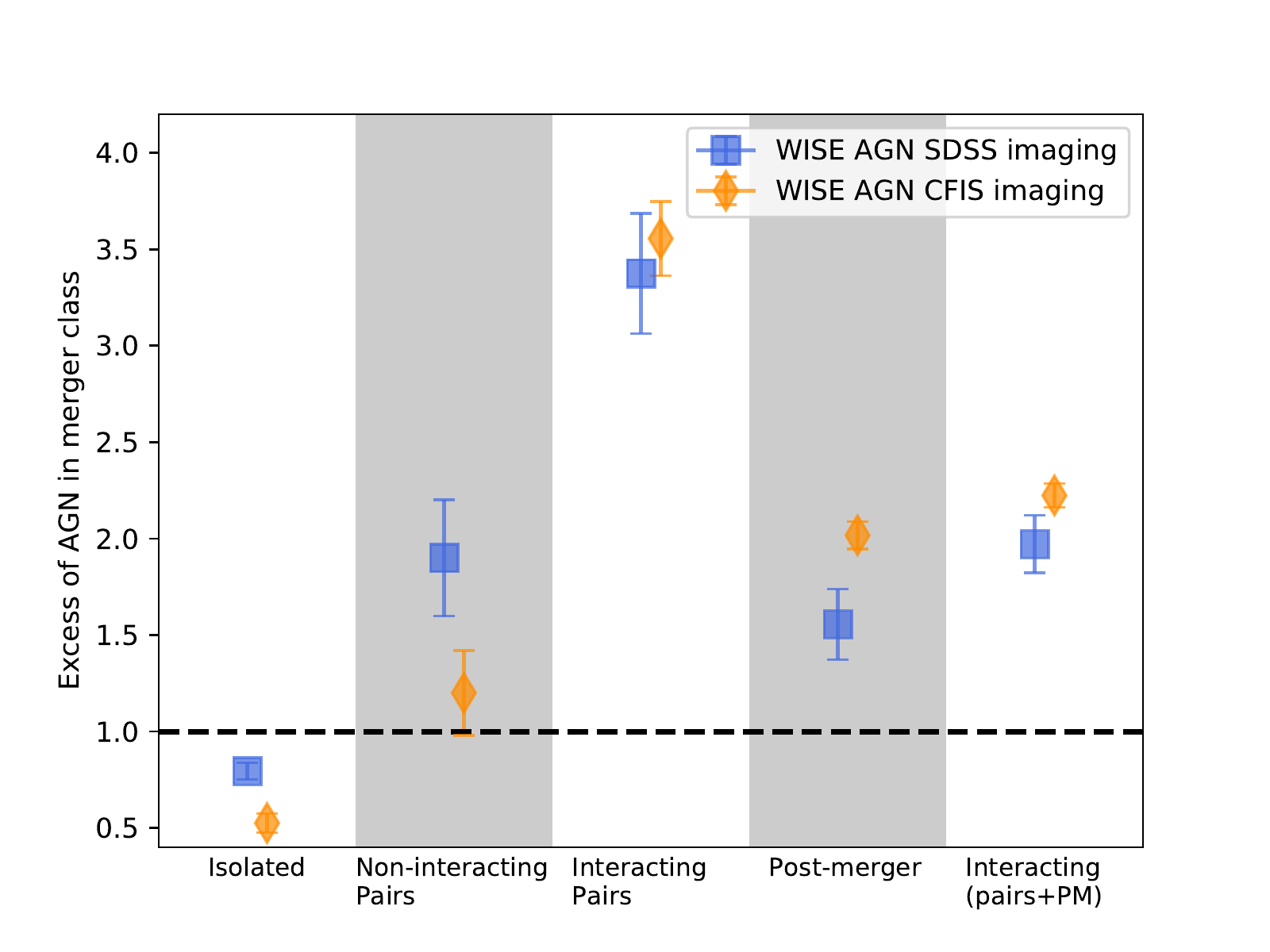}
        \caption{The excess of WISE selected AGN in each morphology class, i.e. the number of AGN
          divided by the number of controls in each category.  Errors are the standard errors as computed from binomial statistics.  The horizontal dashed
          line shows a ratio of one, as would be expected if AGN and non-AGN are equally
          represented in each morphology class.   The `interacting'
          category in the right-most column includes interacting pairs and post-mergers and
          hence encompasses all galaxies with discernable tidal features/disturbances.
          The orange diamonds show the morphology results from CFIS imaging, and are
          reproduced from Fig. \ref{AGN_ex_fig}.  The blue squares are also for the WISE
          selected AGN sample, but for visual classifications based on the SDSS $r$-band
          imaging.  Despite the shallower depth and poorer spatial resolution, the excess
          of disturbed morphologies in AGN hosts is seen in a similar level in the SDSS imaging as
        the CFIS imaging.}
    \label{AGN_ex_sdss_fig}
\end{figure}

In Fig. \ref{AGN_ex_sdss_fig} we compare the relative frequencies of
AGN and controls in different morphology classes.  The orange diamonds
show the same WISE AGN fractions in the CFIS data as shown before in
Fig. \ref{AGN_ex_fig}.  In addition, the blue squares
show the results for the same galaxy sample in SDSS imaging.
Despite the very different absolute fractions of morphologically
  disturbed galaxies in the SDSS and CFIS images (Fig. \ref{wise_sdss_AGN_frac_fig}),
  the excess of interacting pairs and
post-mergers in Fig. \ref{AGN_ex_sdss_fig} is broadly similar between the two surveys. Only the post-mergers
(characterized predominantly by faint shells) seem to show any significant
drop in frequency compared with controls in the SDSS imaging.  However, the final column
of Fig \ref{AGN_ex_sdss_fig} indicates that the relatively poor SDSS
imaging would not have prevented us from concluding a factor of 2 excess in
interacting hosts for AGN.

Although the above experiment indicates that image quality may
not explain the lack of interacting galaxies hosting AGN in high $z$
observations, there may be real redshift evolution effects.
Since gas flows in interacting galaxies are sensitive to
their internal properties, such as bulge and gas fractions (e.g.
Di Matteo et al. 2007; Cox et al. 2008; Fensch et al. 2017;
Blumenthal \& Barnes 2018), it is plausible
that the efficiency of inflow evolves with redshift, as these
properties are also evolving.  A further factor may be that
high $z$ studies typically select AGN at X-ray wavelengths.
We have performed our study using optical and mid-IR AGN selection only,
since a suitable X-ray dataset is not available. A complete assessment of the redshift
evolution of the merger-AGN connection and impact of X-ray selection is beyond the scope of
this paper (whose data are at $z \sim 0$), but we conclude that
this remains a possible explanation for observational results in
the literature.

Another point of conflict in the literature is whether or not the fraction of
  merger triggered AGN increases with AGN luminosity.  The results presented in the upper panel of
  Fig. \ref{AGN_ex_LO3_fig} support such a correlation, indicating that higher accretion rates
  may be preferentially afforded by interactions.
  Lower luminosity AGN may still be triggered by interactions, but may have an increasing
  contribution from other more `gentle' processes.  Indeed, Ellison, Patton \& Hickox
  et al. (2015) have shown
  that the radiatively inefficient (low excitation) radio galaxies seem to be triggered
  entirely by secular processes.  Although some other works (e.g. Treister
  et al. 2012) have also found a dependence of merger fraction on AGN luminosity, our
  results are compiled from a single dataset and methodology, making the interpretation
  less susceptible to some selection biases.  Nonetheless, the dependence of merger fraction
  on AGN luminosity is highly controversial.  For example, several works focusing specifically
  on high luminosity (or accretion rate) AGN have failed to find an excess of mergers
  (Mechtley et al. 2016; Villforth et al. 2017; Marian et al. 2019).  The reason for this
  conflict remains unclear; it may be associated with the higher redshifts of these other works
  ($z \sim 0.6$ for Villforth et al. 2017 and $z \sim 2$ for Mechtley et al. 2016 and Marian et al. 2019),
  although the simulations of Steinborn et al. 2018 find that a luminosity dependence becomes
  more (not less) pronounced at higher $z$.   Alternatively, it may be an issue of sample size.
  The samples of Mechtley et al. (2016), Villforth et al. (2017) and Marian et al. (2019) are
  all quite small ($\sim$ 20 AGN).  If we bootstrap sample our optical AGN sample taking only 20
  AGN and their matched controls, we recover the factor of approzimately two excess seen for the full sample
  with convincing significance (the typical error on the factor of two excess is 0.18) in each of 1000 iterations.
  We conclude that sample size is not the reason for disagreement with previous studies, and
  again are left with AGN selection method and redshift as possible culprits. 

The statistics in Tables \ref{opt_agn_tab} and \ref{ir_agn_tab} have
demonstrated that AGN are more frequently associated with mergers than
non-AGN, leading to our conclusion that there is a connection between
the two phenomena at low redshift.  Indeed, the results in Figs. \ref{AGN_frac_fig} and,
  even more so, Fig. \ref{wise_AGN_frac_fig}
  indicate that mergers may play a significant role in triggering moderate luminosity
  AGN at $z \sim 0$.  However, the quantification of precisely what
fraction of AGN are merger triggered, and understanding whether all mergers lead to
an AGN remains a challenge.  One obvious limitation, that is
demonstrated with our comparison to SDSS imaging (see also Bottrell
et al. 2019) is that even with the high quality CFIS imaging it
is not possible to detect \textit{all} merger features.  Secondly,
as we discussed in the Introduction, AGN manifest at a range of wavelengths,
and thus a complete assessment of the merger-AGN connection requires
a panchromatic census of the accretion phenomena.  Whilst both of
these current limitations are, in theory, tractable with improved
observations and large multi-wavelength samples, a further impediment
that is harder to address, is that of timescales.

If the observability timescale of the merger features and the AGN
lifetimes were the same, the fractions of AGN/non-AGN in mergers/non-mergers
would allow a straightforward calculation of the fraction of
AGN triggered by mergers, as well as the fraction of mergers that
trigger AGN.  However, AGN lifetimes may be orders of magnitude
shorter than the merger timescale, 10$^5$ years (e.g. Schawinski et al. 2015),
compared with hundreds of Myr.  Therefore even if \textit{all}
mergers lead to an AGN at some point during the interaction,
that AGN may not be actively accreting at the time of observation.
These complicating factors mean that an accurate quantification
of the fraction of AGN that are triggered by mergers requires both
improved data, as well as assumptions about the merger observability
and AGN timescales, which will always be inherently uncertain.

We conclude this discussion with a reminder that numerous non-merger
origins have also been proposed for fuelling AGN accretion, including
bars, stellar winds, disk instabilities and halo cooling (e.g. Hopkins \& Hernquist 2006;
Menci et al. 2014; Robichaud et al. 2017).  Ultimately, the black hole is not fussy
about where its meal is coming from, hence it seems likely that
mergers are one of several mechanisms that provide effective delivery.
In this paper, we have demonstrated that, regardless of whether one
looks for a merger excess in a sample of AGN, or an excess of
AGN in a merger sample (e.g. Ellison et al. 2011, 2013), there is
a clear AGN-merger connection at low $z$.  However, our results are not
inconsistent with other observational works that have inferred
links to other phenomena (e.g. Bournaud et al. 2012;
Cisternas et al. 2015; Ellison et al. 2015).  

\section{Summary}\label{sec_summary}

In previous work, we have shown that galaxy mergers in the SDSS show
elevated fractions of both optical and mid-IR selected AGN
(Ellison et al. 2011; Ellison et al. 2013; Satyapal
et al. 2013; Khabiboulline et al. 2014; Ellison, Patton \& Hickox 2015).
However, other work that has quantified the morphologies of (higher redshift,
X-ray selected) AGN has failed to find an enhanced fraction of disturbed galaxies
(e.g. Gabor et al. 2009; Cisternas et al. 2011; Kocevski et al. 2012;
Metchley et al. 2016; Hewlett et al. 2017).
These results have been previously reconciled by considering the two different experimental
approaches that have been used.  The first experiment measures an excess frequency
of AGN in a merger sample, demonstrating that interactions
\textit{can} trigger AGN.  The second experiment finds no enhanced
morphological disturbances in an AGN sample, indicating that the \textit{majority}
of AGN are not initiated by mergers.
However, the different redshift regimes, AGN selection techniques and
AGN luminosities potentially undermines attempts to synthesize the various results
into a single physical picture.  To fully understand the impact of the two
experimental approaches on the statistical connection between mergers and
AGN requires that both experiments be conducted on the same sample.

In this paper, we select low $z$ AGN from the SDSS using the same optical and
mid-IR selection as our previous work and quantify the frequency of companions
and disturbed morphologies in deep $r$-band imaging from the Canada-France Imaging
Survey.    Combined with our previous work, this study represents a thorough
assessment of the merger-AGN connection at $z \sim 0$.

\medskip

Our main conclusions are as follows:

\begin{itemize}

\item 63 per cent of optically selected AGN show no compelling signs
  of visual disturbance, nor have a companion within a projected separation
  of 30 kpc (Fig \ref{AGN_frac_fig}), indicating that \textit{the majority of optically
    selected AGN at $z \sim 0$ are not triggered by a recent interaction.}\\

\item  In contrast, at least 59 per cent of mid-IR selected AGN are visually
  disturbed in CFIS $r$-band images (Fig \ref{wise_AGN_frac_fig}) indicating that \textit{galaxy
    interactions may be the dominant trigger for $z \sim 0$ mid-IR selected AGN.}\\
  
\item Compared to a stellar mass and redshift matched control sample of non-AGN
  host galaxies, both the optically selected and mid-IR selected AGN sample
  show an increased frequency of both interacting pairs and post-mergers (Fig \ref{AGN_ex_fig}).
  \textit{Low redshift AGN are (at least) twice as likely as non-AGN to show features of a recent
    interaction}.\\

\item The excess of disturbed morphologies is enhanced for mid-IR selected
  AGN compared to optically selected AGN  (Fig \ref{AGN_ex_fig}), indicating that \textit{obscured AGN
are preferentially associated with merger triggering.} \\
  
\item To investigate the effect of image quality, the morphological
  classification is repeated for the mid-IR sample using SDSS imaging,
  which is 3 mag arcsec$^{-2}$ shallower than CFIS.
  Although the disturbed fraction of mid-IR AGN drops to 25 percent,
  a similar drop in sensitivity to merger features in the control sample
  means that \textit{the factor of two excess in disturbed morphologies
    for mid-IR AGN is still seen in SDSS imaging (Fig. \ref{AGN_ex_sdss_fig}).}\\
  
\item The fraction of optical AGN that show features of an interaction increases with
  the AGN luminosity (as measured by the [OIII] line luminosity, upper panel of Fig \ref{AGN_ex_LO3_fig}),
  \textit{consistent with a scenario in which higher luminosity AGN are more
    frequently triggered by mergers.}\\

\item The fraction of optical AGN that show features of an interaction increases with
  the host galaxy mass (lower panel of Fig \ref{AGN_ex_LO3_fig}) \textit{consistent with a scenario in
    which merger-triggered nuclear accretion occurs more readily at higher masses.}
  We speculate that this may be due to higher gas fractions in lower stellar mass
  galaxies.\\
  
\end{itemize}

Combined with our previous work, we have now demonstrated that the merger-AGN
connection is evident at low $z$ for experiments that both assess the AGN
fraction in mergers, as well as the merger fraction amongst AGN.  We conclude
that there is definitive evidence for a connection between mergers and AGN
triggering at low $z$ for optical and mid-IR selected AGN.

\section*{Acknowledgements}

SLE and DRP gratefully acknowledge support from an NSERC Discovery Grant.  AV gratefully
acknowledges the receipt of a Mitacs Globalink Award, hosted at the University
of Victoria, which funded her contribution to this work.  AV thanks
Trystyn Berg, Maan Hani, Guillaume Thomas and Mallory Thorp for help in the preliminary stages of this project.

This work is based on data obtained as part of the Canada France
Imaging Survey, a CFHT large program of the National
Research Council of Canada and the French Centre
National de la Recherche Scientifique. Based on observations
obtained with MegaPrime/MegaCam, a joint project
of CFHT and CEA Saclay, at the Canada-France-Hawaii
Telescope (CFHT) which is operated by the National Research
Council (NRC) of Canada, the Institut National des
Science de l’Univers (INSU) of the Centre National de la
Recherche Scientifique (CNRS) of France, and the University
of Hawaii. This research used the facilities of the Canadian
Astronomy Data Centre operated by the National Research
Council of Canada with the support of the Canadian
Space Agency.

Funding for the SDSS and SDSS-II has been provided by the Alfred P. Sloan Foundation, the Participating Institutions, the National Science Foundation, the U.S. Department of Energy, the National Aeronautics and Space Administration, the Japanese Monbukagakusho, the Max Planck Society, and the Higher Education Funding Council for England. The SDSS Web Site is http://www.sdss.org/.

The SDSS is managed by the Astrophysical Research Consortium for the Participating Institutions. The Participating Institutions are the American Museum of Natural History, Astrophysical Institute Potsdam, University of Basel, University of Cambridge, Case Western Reserve University, University of Chicago, Drexel University, Fermilab, the Institute for Advanced Study, the Japan Participation Group, Johns Hopkins University, the Joint Institute for Nuclear Astrophysics, the Kavli Institute for Particle Astrophysics and Cosmology, the Korean Scientist Group, the Chinese Academy of Sciences (LAMOST), Los Alamos National Laboratory, the Max-Planck-Institute for Astronomy (MPIA), the Max-Planck-Institute for Astrophysics (MPA), New Mexico State University, Ohio State University, University of Pittsburgh, University of Portsmouth, Princeton University, the United States Naval Observatory, and the University of Washington.

\appendix

\section{Further CFIS image examples}

\begin{figure*}
	\includegraphics[width=18cm]{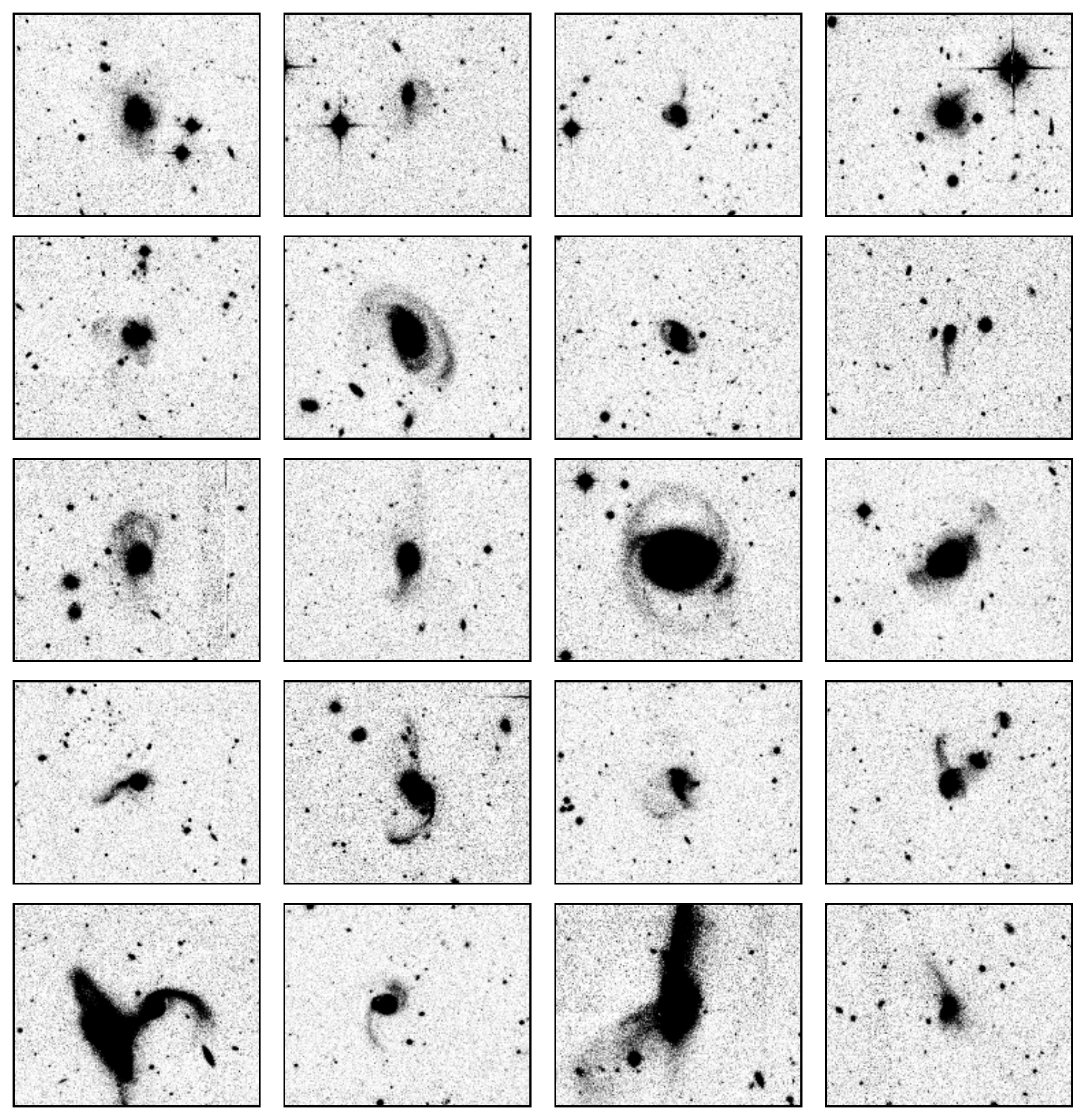}
        \caption{A selection of mergers identified in CFIS imaging.
        All panels are 2 by 1.5 arcminutes.}
    \label{multi_plots_fig}
\end{figure*}

A further selection of mergers identified in CFIS images is
presented in Fig. \ref{multi_plots_fig}, in order to appreciate
the image quality of the survey.

\end{document}